\documentclass[aps,prd,superscriptaddress,twocolumn,longbibliography]{revtex4-1}
\usepackage{amsmath}
\usepackage{amsfonts}
\usepackage{graphicx}
\usepackage{color}
\usepackage{dsfont}
\usepackage{comment}
\usepackage{accents}
\usepackage[section]{placeins}

\begin{document}

\newcommand{\Sb}{ {\bf S} }
\newcommand{\beq}{\begin{equation}}
\newcommand{\eeq}{\end{equation}}
\newcommand{\beqa}{\begin{eqnarray}}
\newcommand{\eeqa}{\end{eqnarray}}
\newcommand{\ben}{\begin{enumerate}}
\newcommand{\een}{\end{enumerate}}
\newcommand{\hs}{\hspace{1.5mm}}
\newcommand{\vs}{\vspace{0.5cm}}
\newcommand{\note}[1]{{\color{red} \bf [#1]}}
\newcommand{\ket}[1]{|#1 \rangle}
\newcommand{\bra}[1]{\langle #1|}
\newcommand{\thickbar}[1]{\accentset{\rule{.4em}{.8pt}}{#1}}
\newcommand{\im}{\dot{\imath}}
\newcommand{\plaq}[4]{
\begin{array}{cc} 
#1 & #2 \\ 
#3 & #4
\end{array}
}
\newcommand{\plaqa}[9]
{
\begin{array}{ccc} 
#1 & #2 & #3\\ 
#4 & #5 & #6\\
#7 & #8 & #9
\end{array}
}

\newcommand{\tg}[1]{\textcolor{blue}{#1}}

\setcounter{equation}{0}
\setcounter{figure}{0}
\setcounter{table}{0}
\setcounter{page}{1}
\makeatletter

\title{Multipole conservation laws and subdiffusion in any dimension}
\author{Jason Iaconis} 
\author{Andrew Lucas}
\author{Rahul Nandkishore}
\affiliation{Department of Physics and Center for Theory of Quantum Matter, University of Colorado, Boulder, Colorado 80309, USA}

\begin{abstract}
Subdiffusion is a generic feature of chaotic many-body dynamics with multipole conservation laws and subsystem symmetries.  We numerically study this subdiffusive dynamics,  using quantum automaton random unitary circuits, in a broad range of models including one dimensional models with dipole and quadrupole conservation, two dimensional models with dipole conservation, and two dimensional models with subsystem symmetry on the triangular lattice.   Our results are in complete agreement with recent hydrodynamic predictions for such theories.
\end{abstract}

\date{\today}

\maketitle

\section{Introduction}

Understanding the dynamics of closed quantum systems in the presence of symmetry
is an important problem which has broad implications for our understanding of
thermalization in quantum many-body systems \cite{Rigol, mblarcmp}.
Random quantum circuit models offer a clean platform where such dynamics can be
studied \cite{Nahum1, PhysRevX.8.021014, Nahum3, amos1, amos2, Prosen}. In theories with $U(1)$
symmetry, such quantum circuit models readily reproduce the expected diffusive dynamics predicted by Fick's law \cite{KhemaniVishHuse,Keyserlingk1}. 
Qualitatively new behavior can emerge, however, when one looks at models with
unconventional conservation laws and symmetries. Two important examples of this are: (\emph{1}) systems
where both total charge and the total dipole moment (or even higher multipole
moments) of the charge are conserved, and (\emph{2}) systems with \emph{subsystem
symmetry}, where total charge is independently conserved along intersecting
sub-dimensional sublattices. Such unusual conservation laws are motivated by the dynamics of fracton systems \cite{chamon, haah, fracton1, fracton2, sub, fractonarcmp} but may also be realized in more conventional settings \cite{KN, gromov2020}.%Both of these examples arise naturally
%when studying the constrained dynamics of \emph{fractons}, the emergent
%elementary excitations of models whose effective low energy description is a
%symmetric tensor gauge theory. 

Implementation of these types of symmetries in local random circuits has
lead to the discovery of unique dynamical phenomena, such as the existence of
charge localization \cite{pai2018localization,Sala_2020, KN} in certain (exponentially large) subspaces, and anomalous
subdiffusion outside them \cite{IaconisVijay}.  
In [\onlinecite{gromov2020}], a hydrodynamic theory was
formulated, which describes the late time and long wavelength behavior of
the charge density.  This hydrodynamic theory is valid in almost all states, outside of the localized subspace (whose measure is exponentially small in system size for ``generic" models).   These analytic predictions were numerically confirmed  in \cite{feldmeier2020,KH}, and analytically in certain large-$N$ models \cite{Pengfei, Ganesan}.
%These analytic predictions were numerically confirmed  in \cite{feldmeier2020}, using the numerical method from \cite{IaconisVijay}, and also in \cite{KH}, using a similar but not identical method, for systems in one dimension. 

In this paper, we numerically test the analytic predictions of \cite{gromov2020} in a much larger family of models. 
In particular, we study 1D circuits with both dipole and quadrupole conservation, 2D square lattice circuits with dipole conservation, and 2D
triangular lattice circuits with non-orthogonal subsystem symmetry.  Throughout we use the numerical method from \cite{IaconisVijay}, which allows us to numerically simulate the
dynamics for very large system sizes and very long circuit depths. We probe the dynamics through the dynamical evolution of wave
functions with special initial charge configurations. This allows us to excite
isolated long wavelength modes so that we can directly
compare with the field theoretic results of [\onlinecite{gromov2020}]. In every case, the numerical results are in agreement with analytic expectations. 
%Furthermore, we study systems in two spatial dimensions. We consider a 2D
%square lattice circuit model with dipole conservation, where we are simulate the
%dynamics of long wavelength charge modes and find that modes at wave number
%$\vec{k} = (k_x,k_y)$ relax at a rate governed by the tensor structure
%$D_{ijkl}k_ik_jk_kk_l$. We also consider a 2D triangular
%lattice circuits where subsystem symmetry where charge is conserved
%independently along the three lattice directions. This extends the previous analysis of
%[\onlinecite{IaconisVijay}] to non-orthogonal subsystem symmetries,
%and we observe new behavior unique to this case. 

\section{Automaton Dynamics}
The numerical method we employ is based on quantum automaton circuits \cite{GopalakrishnanBahti, Alba, IaconisVijay}.  Cellular automaton dynamics are defined as any unitary evolution which does not
generate entanglement in an appropriate basis. Under an automaton gate, $U$, an
initial computational basis state $\ket{m}$ becomes
\begin{eqnarray}
U \ket{m} = e^{i \theta_m} \ket{\pi(m)}. \label{eqn:BasisState}
\end{eqnarray}
where $\pi \in S_D$ is an element of the permutation group on the $D$ elements
which form the computational basis states.
If we start with an arbitrary initial state
\begin{eqnarray}
\ket{\psi_0} = \sum_m c_m \ket{m},
\end{eqnarray}
we have
\begin{eqnarray}
U \ket{\psi_0} = \sum_m c_m e^{i \theta_m} \ket{\pi(m)}
\end{eqnarray}
For computational basis states, $\ket{m}$, no entanglement is generated by the
dynamics and we can therefore exactly track the evolution of $\ket{m(t)}$. 
While the evolution of a single computational basis state $\ket{m}$ is fully
classical, it is important to note that automaton circuits do generate volume
law entanglement when acting on product states which are \emph{not} initially in
the computational basis.  Further, as explained extensively in
[\onlinecite{IaconisVijay}],
automaton dynamics generically generate volume law operator entanglement, %so
%that
%\begin{eqnarray}
%U^\dagger \mathcal{O} U = \sum_S a_s(t) S,
%\end{eqnarray}
which allows us to numerically study the hydrodynamics of operator spreading. 
In all cases, we are able to simulate the dynamics classically by Monte Carlo sampling
random states $\ket{m}$ and tracking the evolution as in
Eq.~\ref{eqn:BasisState}. When using this to evolve operators or arbitrary
initial states, this protocol amounts to a type of quantum Monte Carlo.  

\begin{figure}
\includegraphics[scale=0.55]{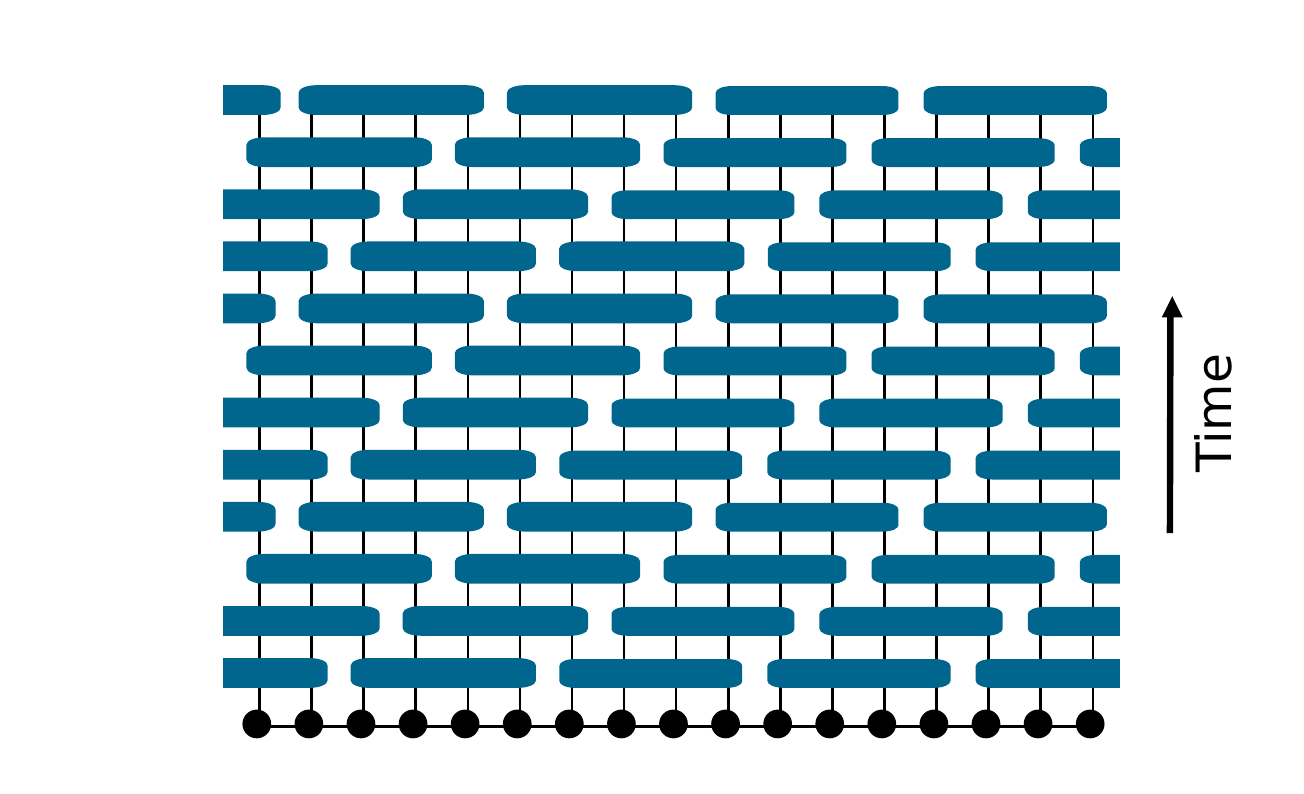}
\caption{The random circuit architecture used for 1D circuits. Each site
contains a 3 state qudit, and the automaton gates of size $|G|$ (shown here for
$|G|=4$), randomly permute the charge configurations in a way which is
consistent with the symmetry contraints. A similar arcitecture is used for 2D
circuits with gates of size $|G|=G_x\times G_y$, with successive layers shifted
by one site in the $x$-direction for $G_x$ layers followed by a single site
shift in the $y$-direction.  Examples of allowed charge permutations are shown
in Fig.~\ref{fig:2dlat}. }
\label{fig:circuit}
\end{figure}

Let us contrast the automaton method with earlier numerical methods, which were largely based instead on simulating Haar random circuits  \cite{KhemaniVishHuse,Keyserlingk1,pai2018localization, Sala_2020, KN}.  Here, one constructs local unitary matrices which are block
diagonal in each symmetry sector, but for which each block is chosen to be Haar
random.  This approach is possible for small systems, and has been performed in
studies with very simple symmetry
constraints.  However, for the constrained dynamics of interest in this paper, Haar random circuits
quickly become intractable for systems whose on-site Hilbert space dimension is larger than 2, and/or which have many more than 10-20 lattice sites.  Moreover,  in
the case of higher-moment conserving dynamics, very large systems are 
required to observe the correct quantitative behavior.  This is why we use the automaton dynamics.

Despite their apparent simplicity, automaton circuits appear
to produce generic chaotic dynamics~\cite{IaconisVijay,iaconis2020quantum}. We expect that essentially any property of a generic
Haar random unitary dynamics can also be seen in a corresponding automaton
circuit, as long as we choose appropriate initial conditions. 

In this work, we will mainly focus on the evolution of spin-1 wave functions
$\ket{\psi_0} = \sum_m c_m \ket{m}$, $\ket{m} \in \{\ket{+},\ket{0},\ket{-}\}$,  
where the coefficients $c_m$ are chosen to induce a net multipole moment. 
We then apply a random local unitary circuit $U = \prod_{i,t} U_{i,t}$, composed
of local unitary gates of size $|G|$ as shown in Fig.~\ref{fig:circuit}. 
The gates $U_{i,t}$ perform a random permutation of the spins within
the block diagonal symmetry sector consistent with the charge and multipole
conservation laws.

\begin{figure}[t]
\hspace{-5mm} {\bf a)} \includegraphics[scale=0.18]{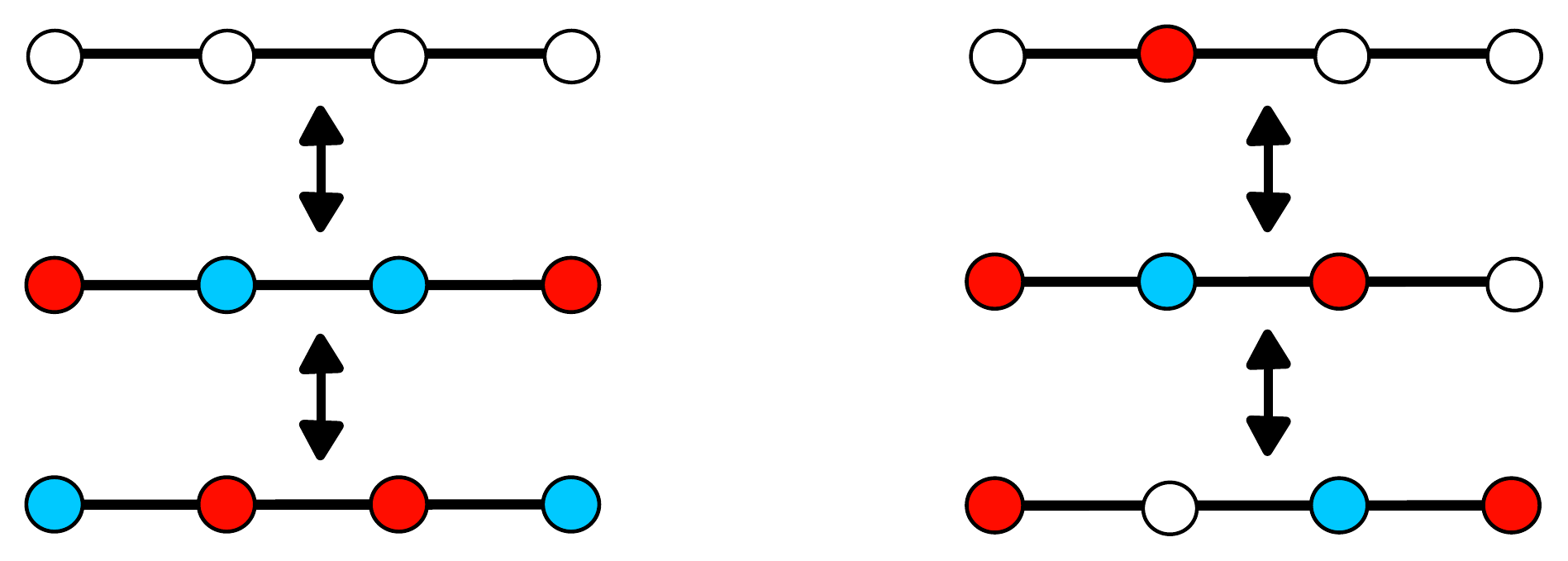} 
\hspace{5mm}  {\bf b)}	\includegraphics[scale=0.18]{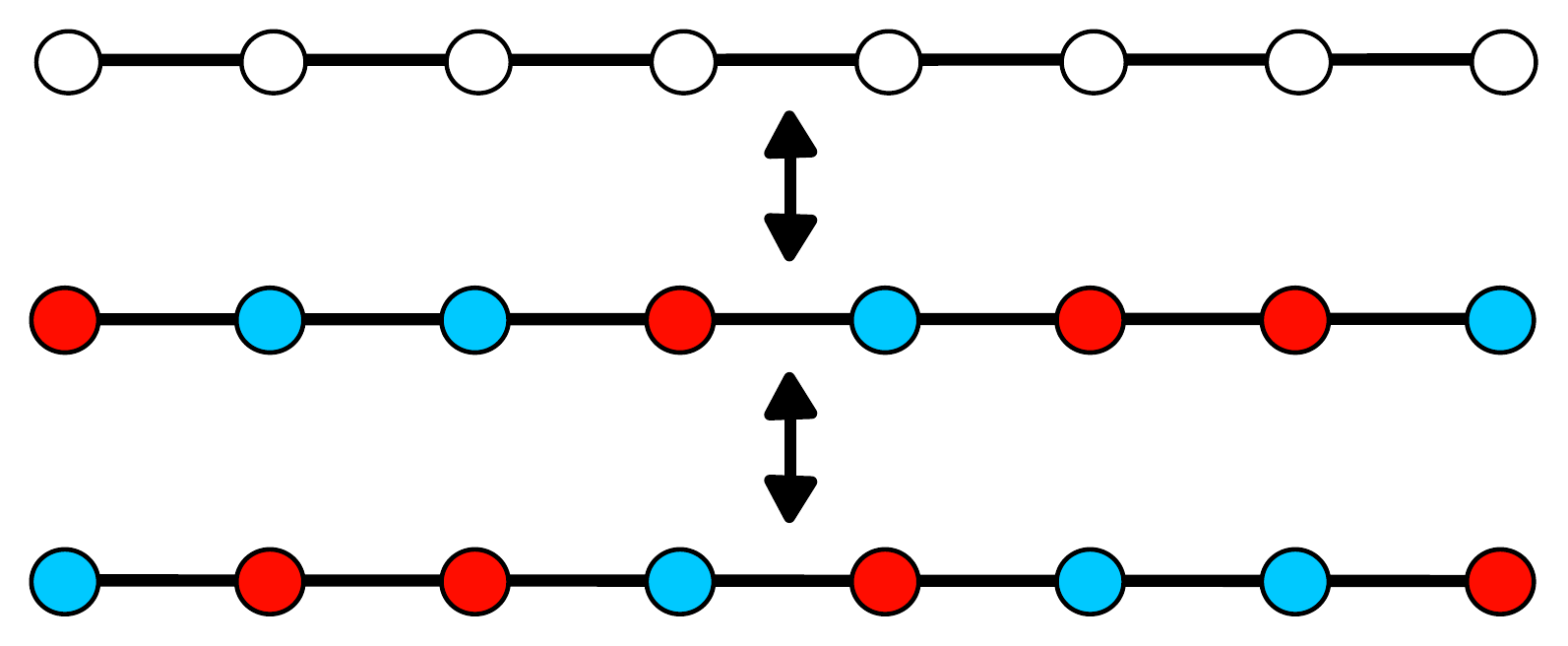} 
\\
\vspace{1.5pc}
\hspace{-82mm} {\bf c)}\\
 \includegraphics[scale=0.35]{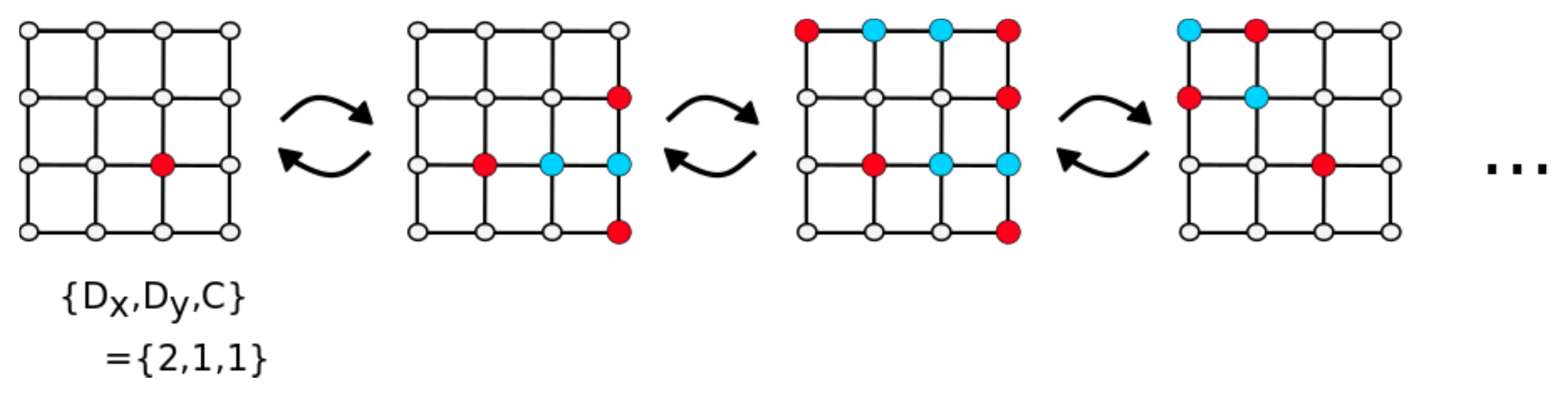}
\caption{Examples of allowed charge permutations for {\bf a)} 4 site 1D gates with dipole
conservation, {\bf b)} 8 site 1D gates with quadrupole conservation and {\bf c)}
2D 4x4 unitary gates with dipole conservation in both the x and y directions.}
\label{fig:2dlat} 
\end{figure}

\section{1D Circuits with Dipole Conservation}

\begin{figure}[!htbp]
\includegraphics[scale=0.33]{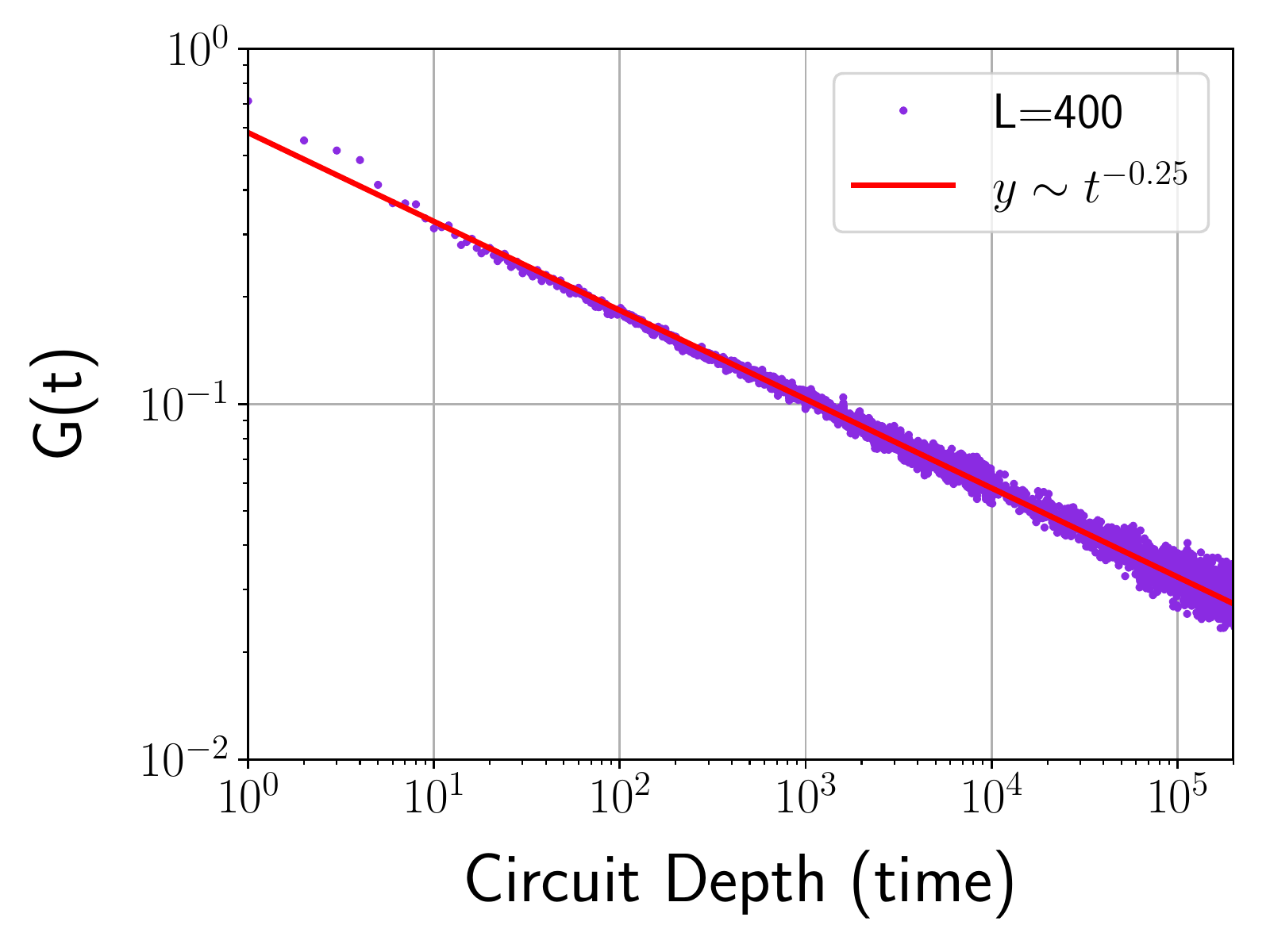}
\caption{ We measure the autocorrelation time for 1D circuits with dipole
conservation. The slow relaxation rate at the long wavelengths gives an
anomalously slow decay of $G(t) = \langle S^z(x,t)S^z(x,0) \rangle$. In this case we
see that $G(t) \sim t^{-0.25}$.} 
\label{fig:dipole2}
\end{figure}

We first simulate 1D circuits with dipole conservation laws using the method of \cite{IaconisVijay}. This
is an extension of the numerical analysis of \cite{feldmeier2020}, where the autocorrelation
function $\langle S^z(x,t)S^z(x,0) \rangle$ was studied and was found to decay sub-diffusively. 
In this work, we also present the Fourier correlators $\langle S^z(k,t)S^z(k,0)\rangle$, and explicitly demonstrate the exponential decay of hydrodynamic modes predicted by \cite{gromov2020}.  The local charge density in dipole conserving systems obeys
the equation of motion \cite{genem, gromov2020}
\begin{eqnarray}
\partial_t \rho + \partial_i \partial_j J_{ij} = 0 . \label{eqn:eom}
\end{eqnarray}
Hydrodynamics implies that in one dimension \begin{equation}
J_{xx} = B_1 \partial_x^2 \rho . \label{eq:Jij}
\end{equation} The decay rate of the Fourier correlator is then \begin{equation}
C(k,t) = \langle S^z(k,t)S^z(k,0)\rangle \sim \exp[-Bk^4t]. \label{eq:Ckt}
\end{equation}  
This then implies the autocorrelator
\begin{eqnarray}
G(t) = \langle S^z_i(x,t) S^z_i(x,0) \rangle \sim t^{-1/4}.
\end{eqnarray}
In Fig.~\ref{fig:dipole2}, as in \cite{feldmeier2020}, we see that the decay indeed follows this
scaling form very closely.  

We now would like to more directly probe the equation of motion
Eq.~\ref{eqn:eom}, by studying the relaxation rate of specific charge modes.
To do this, we simulate the dynamics of the spin-1 state
\begin{eqnarray}
\ket{\psi_0} &=& \frac{1}{\mathcal{N}}\otimes \ket{\vec{h}_i}  \\
\ket{\vec{h}_i} &=& (1+h_i)\ket{+} \, + \, \ket{0} \, + \, (1-h_i)\ket{-} \\
h_i &=&  \left \{ \begin{array}{l} |h| \text{ \quad if  \, $x \le \frac{L}{4}$
or $x \ge \frac{3L}{4}$ }\\ 
-|h| \text{ \quad if  \, $\frac{L}{4} < x < \frac{3L}{4}$}\end{array} \right. ,
\nonumber
\end{eqnarray} 
where $\mathcal{N}$ is a normalization factor. 
That is, we simulate the dynamics of a wave function with zero net charge and
dipole moment, but with a finite quadrupole moment. The charge is configured in
a square wave of the form $\ket{+--+}$, with wavelength $\lambda = L$.

\begin{figure}[t]
{\bf a)}\includegraphics[scale=0.35]{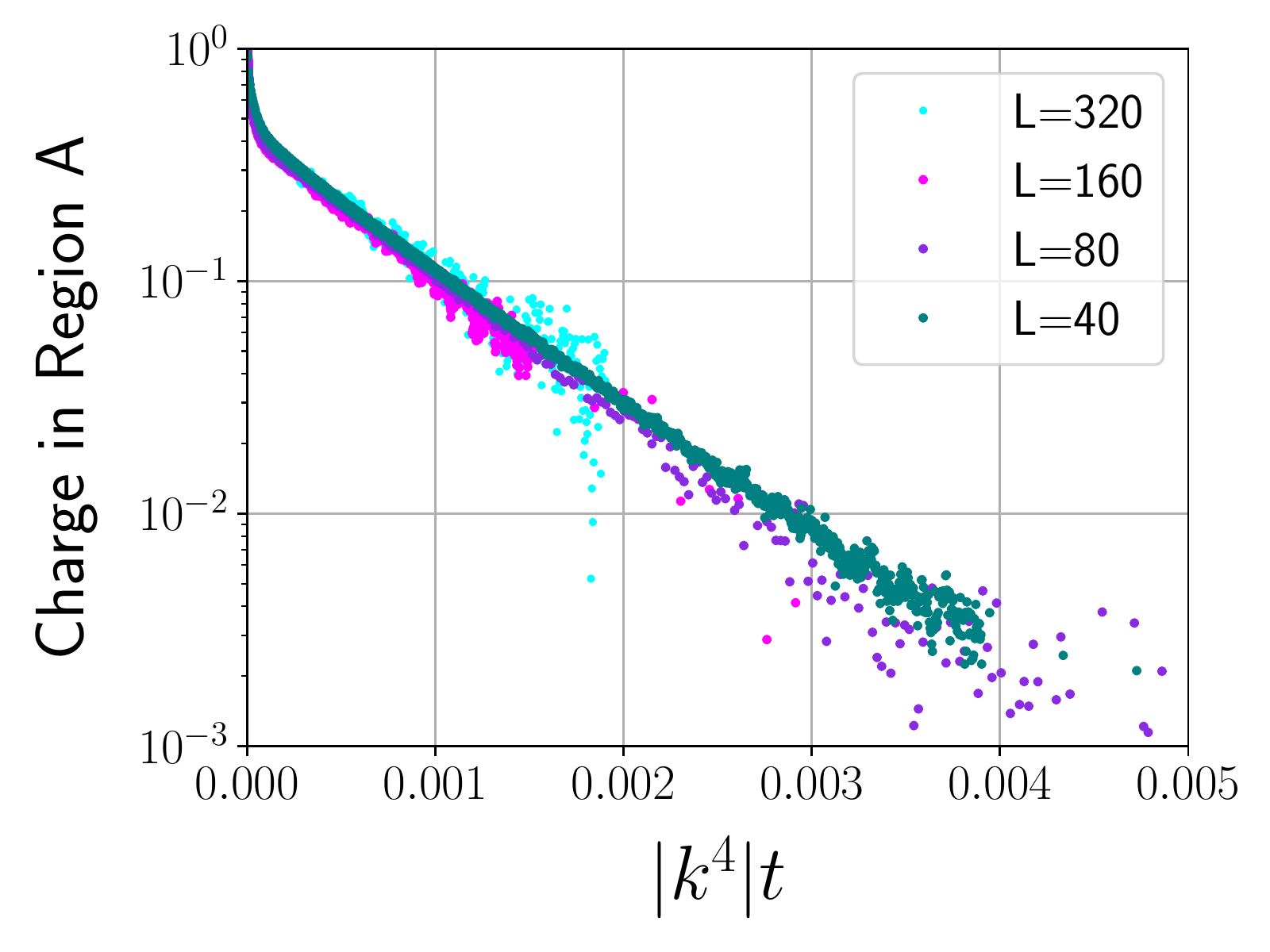} \\
{\bf b)}\includegraphics[scale=0.35]{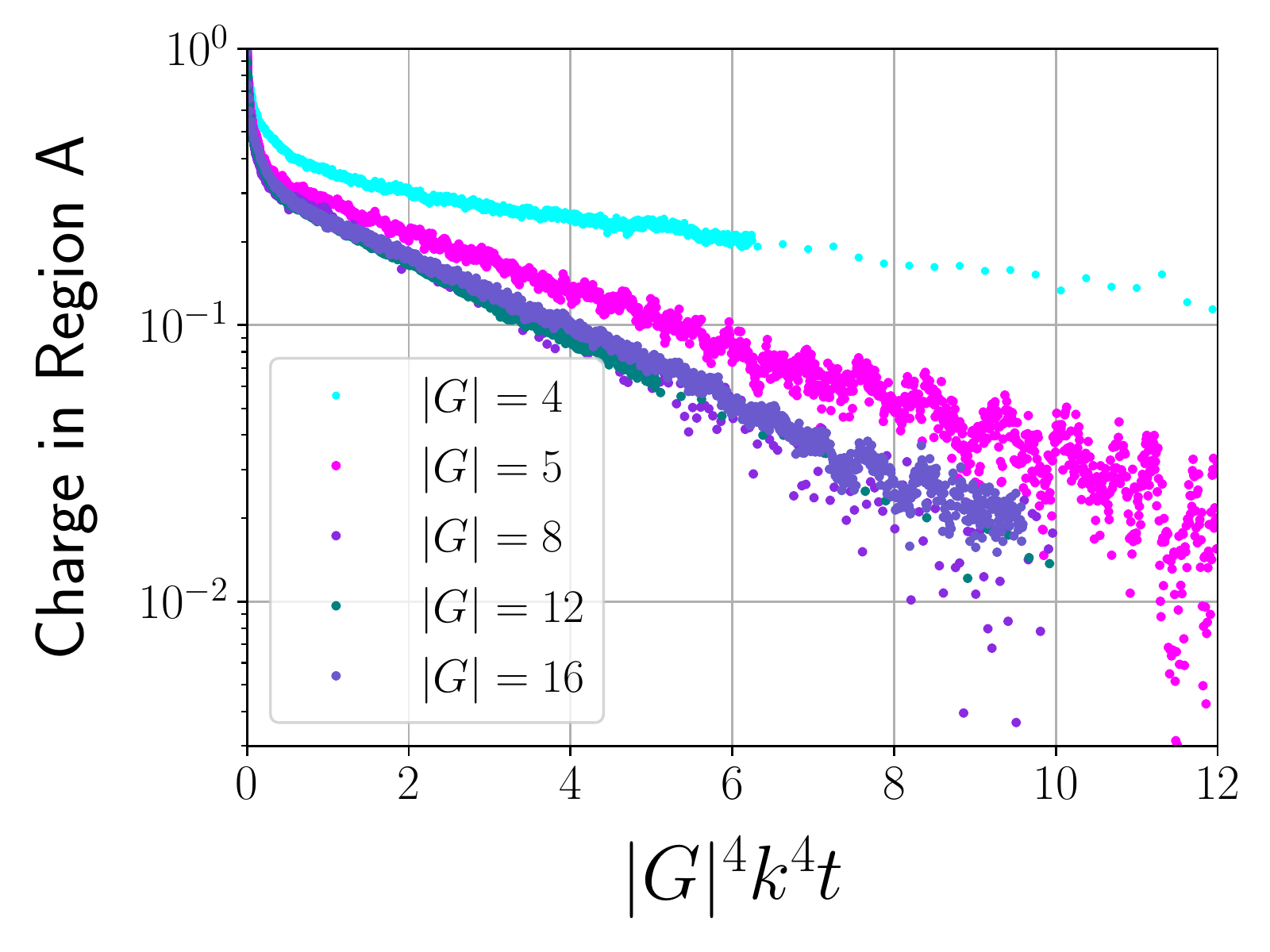}
\caption{$\bf{a)}$ The dissipation time for the 1D dipole conserving circuit
with gates of size $|G|=8$, starting from an initial square wave state with
wavelength $L$. The charge in the region $A = \{x<\frac{L}{4}\}$ decays exponentially with a
relaxation rate $\tau$ which depends on the wavelength $k$ of the initial
state. Note that the time axis is scaled by the system size to the fourth power
$t/L^4$, indicating that the longest wavelength mode relaxes like $\tau \sim
k^4$. {\bf b)} The charge dissipation for the same system as a function of gate
size $|G|$ eventually collapses onto a universal function when the time axis is
scaled like $t (|G|/L)^4$. } \label{fig:dipole1}
\end{figure}

The value of the field $|h|$ is adjusted so that the net charge in each quadrant
of the lattice is the same for different system sizes.    At long times, the state will relax
to a fully neutral state.  
We then measure the net charge in region $A= \{x<\frac{L}{4}\}$ as a function of time, and
therefore measure the relaxation rate for charge modes $k=\frac{2 \pi}{L}$.  We expect the charge to decay exponentially, as in (\ref{eq:Ckt}).
The results are shown in Fig.~\ref{fig:dipole1} a), where we simulated circuits
with gates of size $|G|=8$. We find that all curves
collapse onto the universal function
\begin{eqnarray}
C(t,L) = F\left (\frac{t}{L^4}\right).
\end{eqnarray}
Note that to see this data collapse, we must simulate circuits to a depth, $D$, which
scales with system size like $D \sim L^4$.  For the largest systems we simulate,
we have $L=320$, and go up to depths $D \sim 2.0\times10^7$.  We therefore see
the benefit of performing the simulation using automaton circuits, which allow
us to efficiently study such large system sizes and extreme circuit depths. 

In Fig.~\ref{fig:dipole1} b), we looked at the dependence of the relaxation rate
on gate size.  We know that for sufficiently small gate size, the Hilbert space
for quantum circuits with multipole conservation laws will exhibit `strong shattering'  \cite{KN, Sala_2020} and that the
system will not thermalize at all. For dipole conserving circuits, it is known
that gates of size $|G|=4$ are needed for thermalization to occur.  
In the limit of large gate size we might expect that
\begin{eqnarray}
C(t,L) = F\left[t \left (\frac{|G|}{L}\right)^4 \right ]. 
\end{eqnarray}
Our results confirm this scaling form. The data appears to collapse onto a
universal function for $|G|>5$. For $|G|=4$, the charge still decays, 
however the relaxation rate appears to be far slower. This short wavelength effect 
is likely due to the shattering of the Hilbert space which is more extreme for
smaller gate sizes.

\section{1D Circuits with Quadrupole Conservation}

We now extend these results to 1D circuits with quadrupole conservation laws. 
Much of the analysis remains the same as the case with dipole conservation. The
equation of motion governing the time evolution of the charge density now
includes an additional two factors of the spatial derivative:
\begin{eqnarray}
\partial_t \rho - B \partial_x^6 \rho = 0,
\end{eqnarray}
which in turn implies that density modulations at wave number $k$ relax in time
$\tau \sim k^6/B$.
 
To see this behavior in our lattice model, we now simulate
the evolution of an initial wave function with a net zero quadrupole moment, but
a nonzero long wavelength octopole moment. 
\begin{eqnarray}
\ket{\psi_0} &=& \frac{1}{\mathcal{N}}\otimes \ket{\vec{h}_i}  \\
\ket{\vec{h}_i} &=& (1+h_i)\ket{+} \, + \, \ket{0} \, + \, (1-h_i)\ket{-} \\
h_i &=&  \left \{ \begin{array}{l} |h| \text{ \quad if  \, $x \in A $ }\\ 
-|h| \text{ \quad if  \, $x \in  B $}\end{array} \right. ,
\nonumber
\end{eqnarray} 
where $\mathcal{N}$ is a normalization factor and the regions $A$ and $B$ are
defined as 
\begin{eqnarray}
A &=& \left [0,\frac{L}{8} \right ]
		\cup \left [\frac{3L}{8},\frac{L}{2} \right] 
		\cup \left [\frac{5L}{8},\frac{7L}{8} \right] \\ 
B &=& \left [\frac{L}{8},\frac{3L}{8} \right]
		\cup \left[\frac{L}{2},\frac{5L}{8} \right ] 
		\cup \left [\frac{7L}{8},L \right ].
\end{eqnarray}
We measure the relaxation time in these states, using circuits with gates of
size $|G|=12$.

\begin{figure}[t]
\includegraphics[scale=0.4]{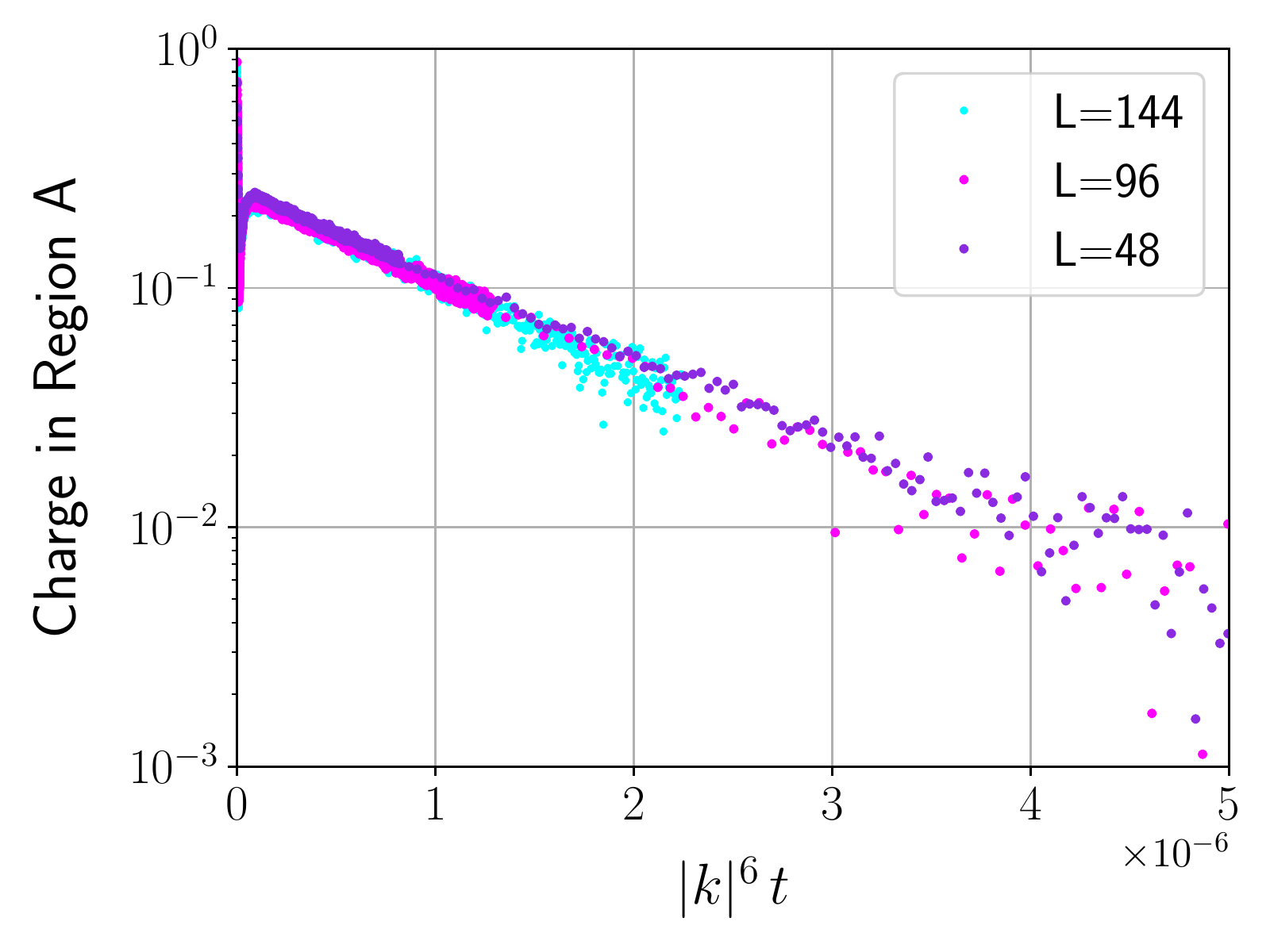}
\caption{Dissipation time for different system sizes in the quadrupole
conserving circuit. Note that all curves collapse onto a universal function of
$\exp(-c t/L^6)$, showing that the relaxation rate scales like $\tau \sim k^6$}
\label{fig:quad1}
\end{figure} 

The results are shown in Fig.~\ref{fig:quad1}. We see that the data collapses
onto the universal function
\begin{eqnarray}
C(t,L) = C(0) F\left( \frac{t}{L^6} \right) .
\end{eqnarray}
and that the decay is exponential in time. This implies that charge relaxes like
$\tau \sim e^{-k^6 t}$.

Again, the system sizes we are able to study are limited by the
circuit depths that can be simulated in a reasonable amount of time. In this
case, the circuit depth must scale like $D\sim L^6$. The largest systems we
simulated contained $L=144$ and again went to depths $D=2\times10^7$.

\section{2D Circuits with Dipole Conservation}

We now turn our attention to studying multipole conservation in 2D systems.
This is a case not studied in previous literature \cite{IaconisVijay,
feldmeier2020}.   The equations (\ref{eqn:eom})-(\ref{eq:Jij}) continue to hold.  However, in two dimensions and on a square lattice, the allowed tensor structures in $J_{ij}$ are non-trivial: \begin{equation}
J_{ij} = -\left[B_1 \delta_{ijkl} + B_2 \delta_{ij}\delta_{kl} + B_3 \delta_{ik}\delta_{jl} \right]\partial_k\partial_l \rho
\end{equation}
where $\delta_{ijkl}=\delta_{ij}\delta_{ik}\delta_{il}$ demands all 4 indices are the same.   Hence we predict that 
\begin{eqnarray}
C(k,t) &=& C(k,0) \exp\left[ - D(k_x,k_y) t \right] \\
D(k_x,k_y) &=& D_{ijk\ell} k_ik_jk_kk_\ell,
\end{eqnarray}
where
\begin{eqnarray}
D_{ijk\ell} k_ik_jk_kk_\ell = (B_2+B_3) (k_x^2+k_y^2)^2 + B_1(k_x^4+k_y^4).
\nonumber \\ 
\label{eqn:tensor}
\end{eqnarray}
Of most interest is the ratio \begin{equation}
b = -\frac{B_1}{B_2+B_3} \label{eqn:fit_anis}
\end{equation}
which encodes the level of anisotropy due to the square lattice.  We can determine $b$ by looking at the
relaxation rate for systems with different aspect ratios $R=L_x/L_y$.
%Note that this case is a more general conservation law than that of a 2D circuit
%with x and y subsystem symmetry which was studied in Ref.~\onlinecite{IaconisVijay}. 
%In that case, not only are the net dipoles $D_x$ and $D_y$ are conserved, but
%separate $U(1)$ charges $C_x$ and $C_y$ must be conserved on all rows and columns of the lattice. 

%For the 2D dipole conserving circuit, the equation of motion for the density now reads
%\begin{eqnarray}
%\partial_t \rho + B (-\nabla)^4 \rho = 0.
%\end{eqnarray}

%The charge relaxation in this case is again expected to behave like like
%$|\vec{k}|^4$. However, we must be somewhat careful to include a nontrivial
%tensor structure which is allowed by the square lattice symmetry. In this case,
%we have that the charge in a region decays like

%
%\begin{figure}[t]
%\hspace{-5mm} a) \includegraphics[scale=0.18]{./figs/1d_gates_dipole.pdf} 
%\hspace{5mm}  b)	\includegraphics[scale=0.18]{./figs/1d_gates_quad.pdf} 
%\\
%\vspace{1.5pc}
%\hspace{-80mm} c)\\
% \includegraphics[scale=0.35]{./figs/2d_4x4lat.pdf}
%\caption{Examples of allowed charge permutations for {\it a)} 4 site 1D gates with dipole
%5conservation, {\it b)} 8 site 1D gates with quadrupole conservation and {\it c}
%2D 4x4 unitary gates with dipole conservation in both the x and y directions.}
%\label{fig:2dlat} 
%\end{figure}

\begin{figure}[t]
\includegraphics[scale=0.40]{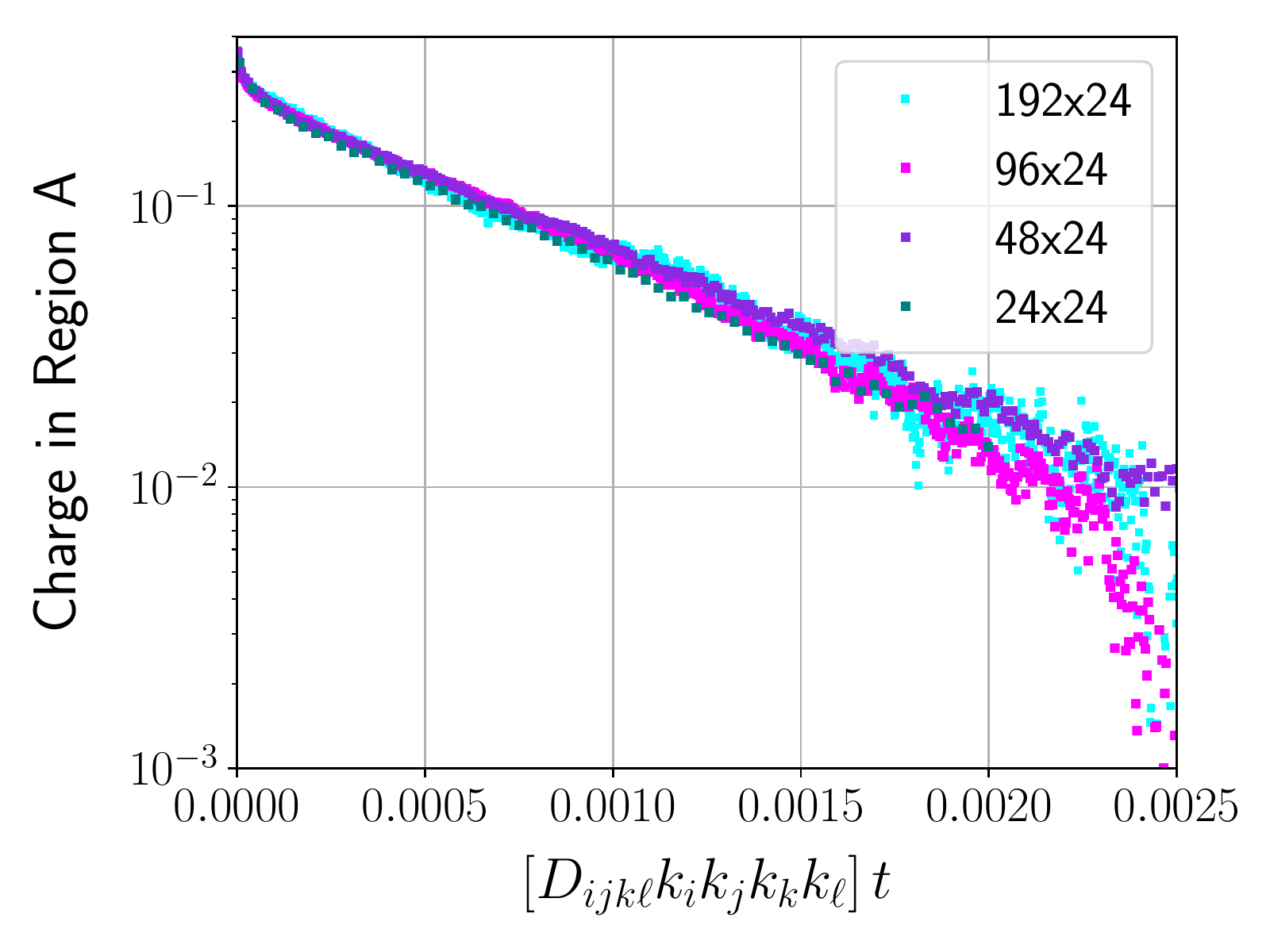}
%\includegraphics[scale=0.40]{./matplotlib/sub2d_1.pdf}
%\caption{Data collapse for the 2D model [IMPROVE]}
\caption{Dissipation time for the 2D dipole conserving model. All curves
collapse onto a universal function $\exp(-D_{ijk\ell}k_ik_jk_k k_\ell \, t)$
with $\{k_x,k_y\} = \{\frac{2\pi}{L_x},\frac{2\pi}{L_y}\}$.  The form of the tensor
$D_{ijk\ell}$ is given in Eq.~\ref{eqn:tensor}.  We see that the relaxation rate
generically scales like $\tau \sim |k|^4$. We simulate the charge relaxation for
systems with different aspect ratios. This allows us to fit the degree of anisotropy
introduced by the square lattice, by fitting with the parameter defined in
Eq.~\ref{eqn:fit_anis}. }
\label{fig:sub2d}
\end{figure}

We simulate the 2D model for various aspect ratios, using gates of size
$|G|=4\times4$. Using these relatively large gate sizes allows the charge to decay
more quickly, easing the computational burden of the simulation since extreme
circuit depths are not needed. Some examples of allowed charge configurations
are shown in Fig.~\ref{fig:2dlat}, whereby a local charge can move by emitting
an $x$ or $y$ dipole. Charge configurations with a net-zero dipole moment can
also be created from the vacuum state within the $4\times 4$ sublattice.

We initialize our wave function to have a net zero dipole moment but a finite
quadrupole moment by dividing the lattice into four quadrants and inducing a
positive (negative) net charge in the lower left and upper right (lower right
and upper left) quadrants.  

The results of our 2D simulation are shown in Fig.~\ref{fig:sub2d}. The optimal
data collapse occurs with $b\sim 0.8$, where $b$ is defined as in
Eq.~\ref{eqn:tensor}.  We find that, in this case, the long wavelength modes
indeed decay like $|\vec{k}|^4$.

\section{Subsystem Symmetry on the Triangular Lattice}

\begin{figure}[t]
\includegraphics[scale=0.30]{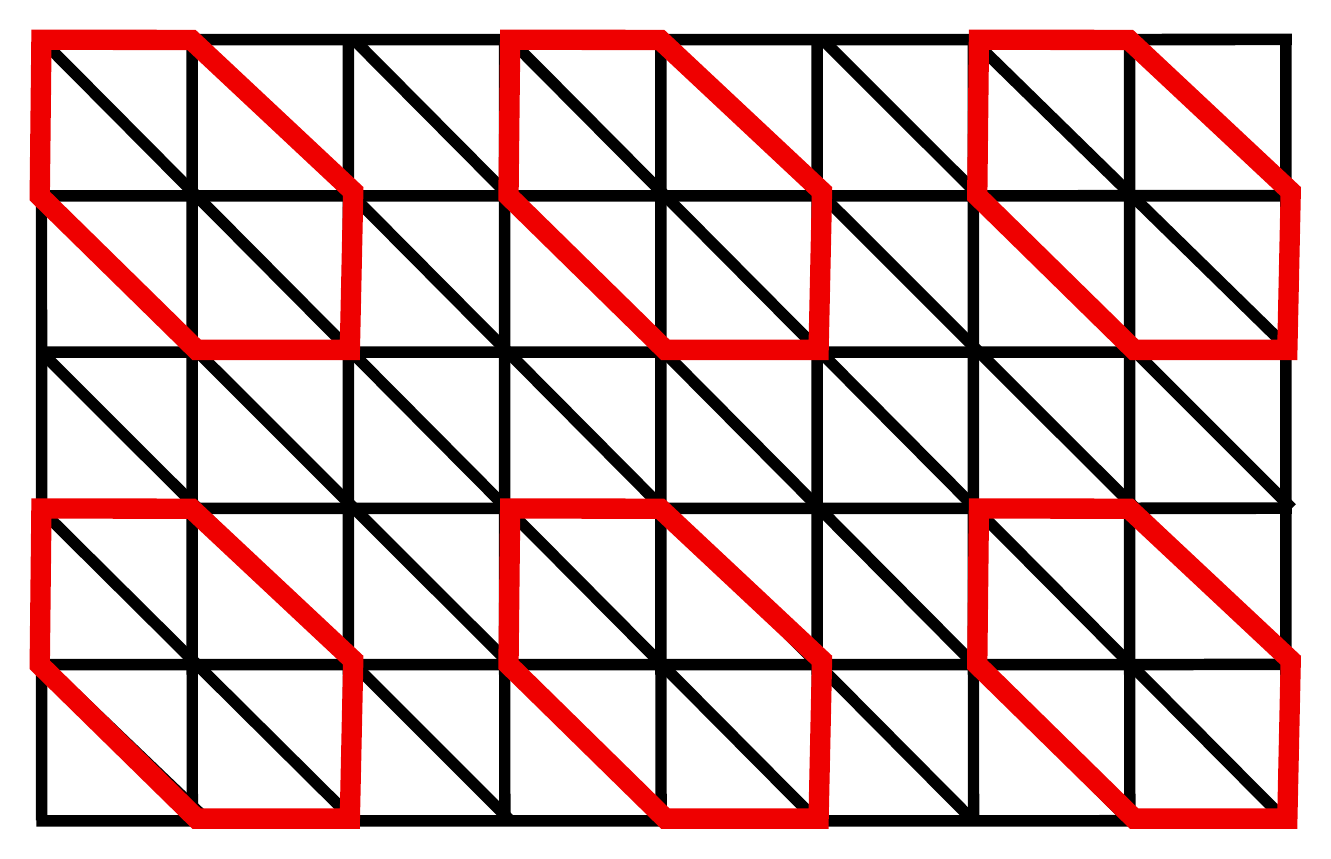}
\hspace{5mm} \includegraphics[scale=0.45]{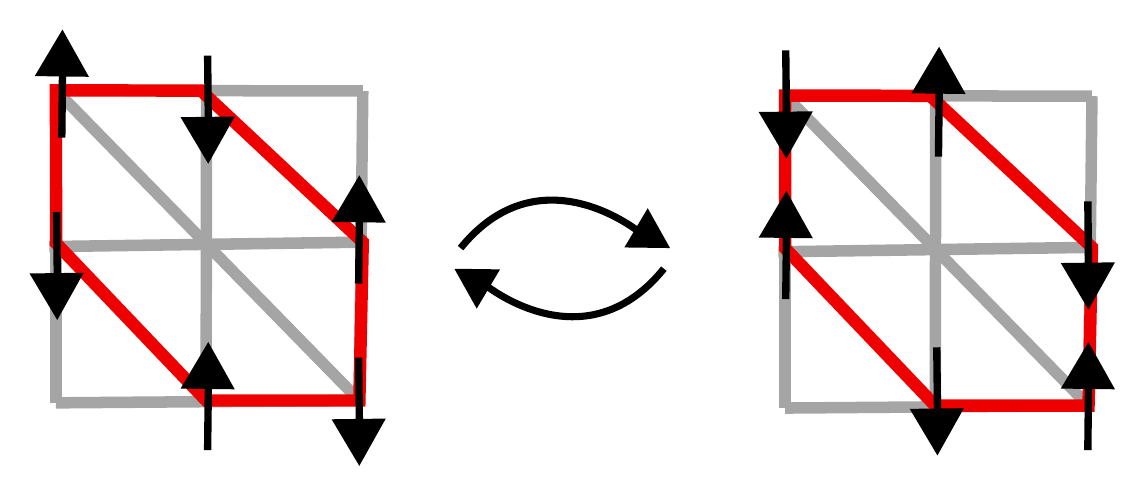}
\caption{The skewed triangular lattice, which is equivalent to the actual
triangular lattice under the discrete circuit dynamics.  The smallest
non-trivial gate which preserves the subsystem symmetry acts nontrivially on the
6 sites of the hexagons shown here. Note that each unit cell consists of 9
sites, but the gates act trivially on 3 of these sites.  The subset of
nontrivial configurations is shown below. Such a tiling
pattern would be repeated and shifted to begin at all sites of the unit cell, so
that one timestep is equal 9 layers of the circuit}
\label{fig:tri1}
\end{figure}

Lastly, we look at a different example of a higher order conservation law:  
 subsystem symmetries.  As mentioned earlier, subsystem symmetry occurs when a
lower dimensional symmetry is embedded in a higher dimensional system. In our
case, we look at systems where charge is conserved on each row of the lattice
individually. On the square lattice and cubic lattice, this was studied in
Ref.~\onlinecite{IaconisVijay}, where subdiffusive behavior was found. Here, we
extend this analysis to the case of \emph{nonorthogonal} subsystem symmetries.
In particular, we look at a triangular lattice system, where charge is conserved
on each row of the triangular lattice.  

In this case, unlike the rest of the paper, we study a spin-1/2 system. The
subsystem symmetry can be implemented in an automaton circuit by applying a gate
which acts on a $3\times3$ sublattice and flips between two specific charge
configurations, illustrated in Fig.~\ref{fig:tri1}.  We then obtain a conserved charge along all lattice directions
$\vec{\lambda}_k$:  for any starting point $\mathbf{x}$ and lattice direction $\mathbf{e}_k$,
\begin{equation}
Q_{k,\mathbf{x}} = \sum_n  S^z_{\mathbf{x}+n\mathbf{e}_k}
\end{equation}
 is conserved.  (Note that these are not all unique charges, as defined above.)  
%\begin{eqnarray}
%C_{k} \equiv \sum_n Z_{n\vec{\lambda}_k} 
%\end{eqnarray} 
%where $Z_{n\vec{\lambda}_k}$ is the Pauli Z operator on site $\vec{r} = n
%\vec{\lambda}_k$.

\begin{figure}[t]
\includegraphics[scale=0.40]{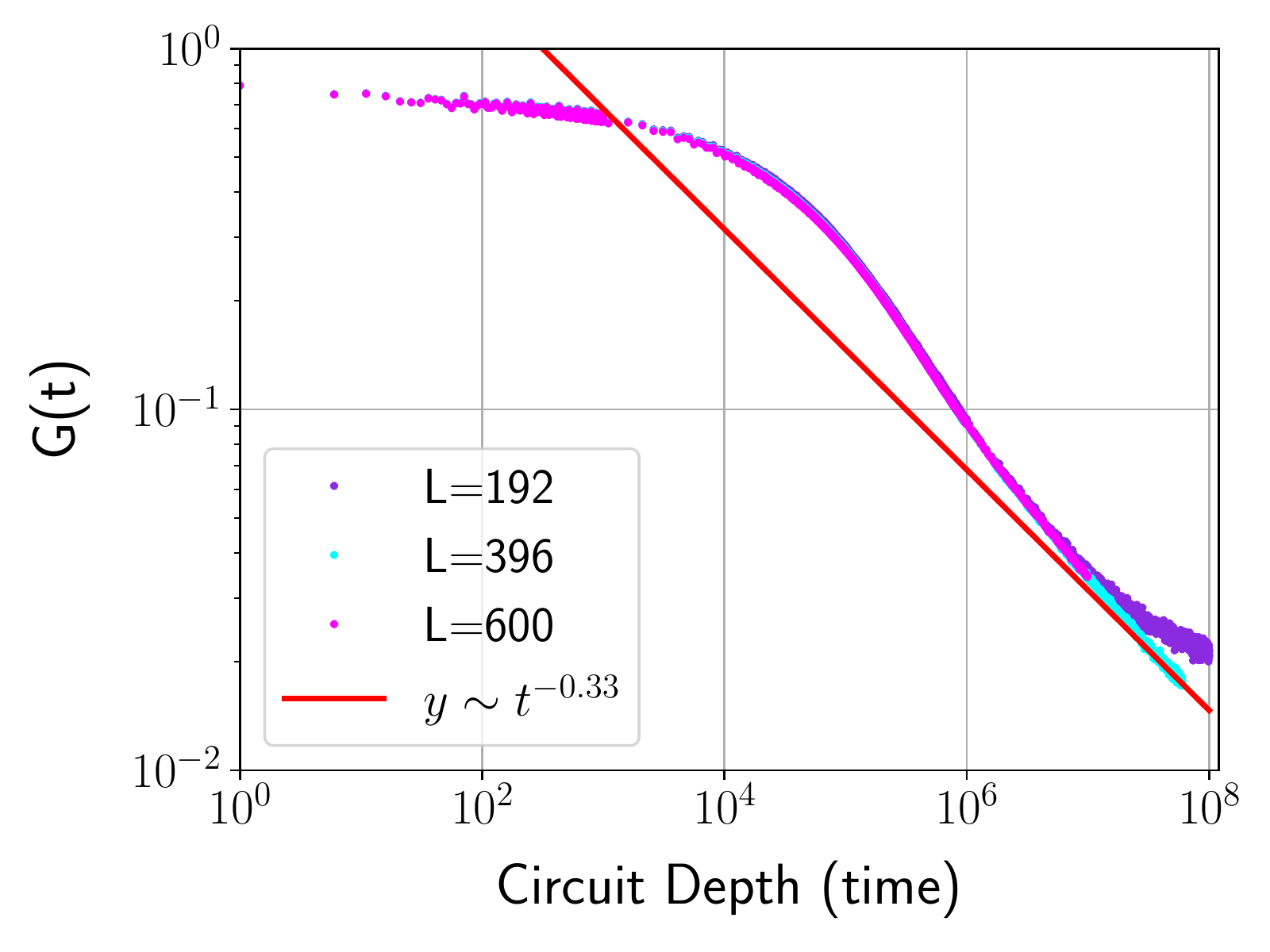}
\includegraphics[scale=0.40]{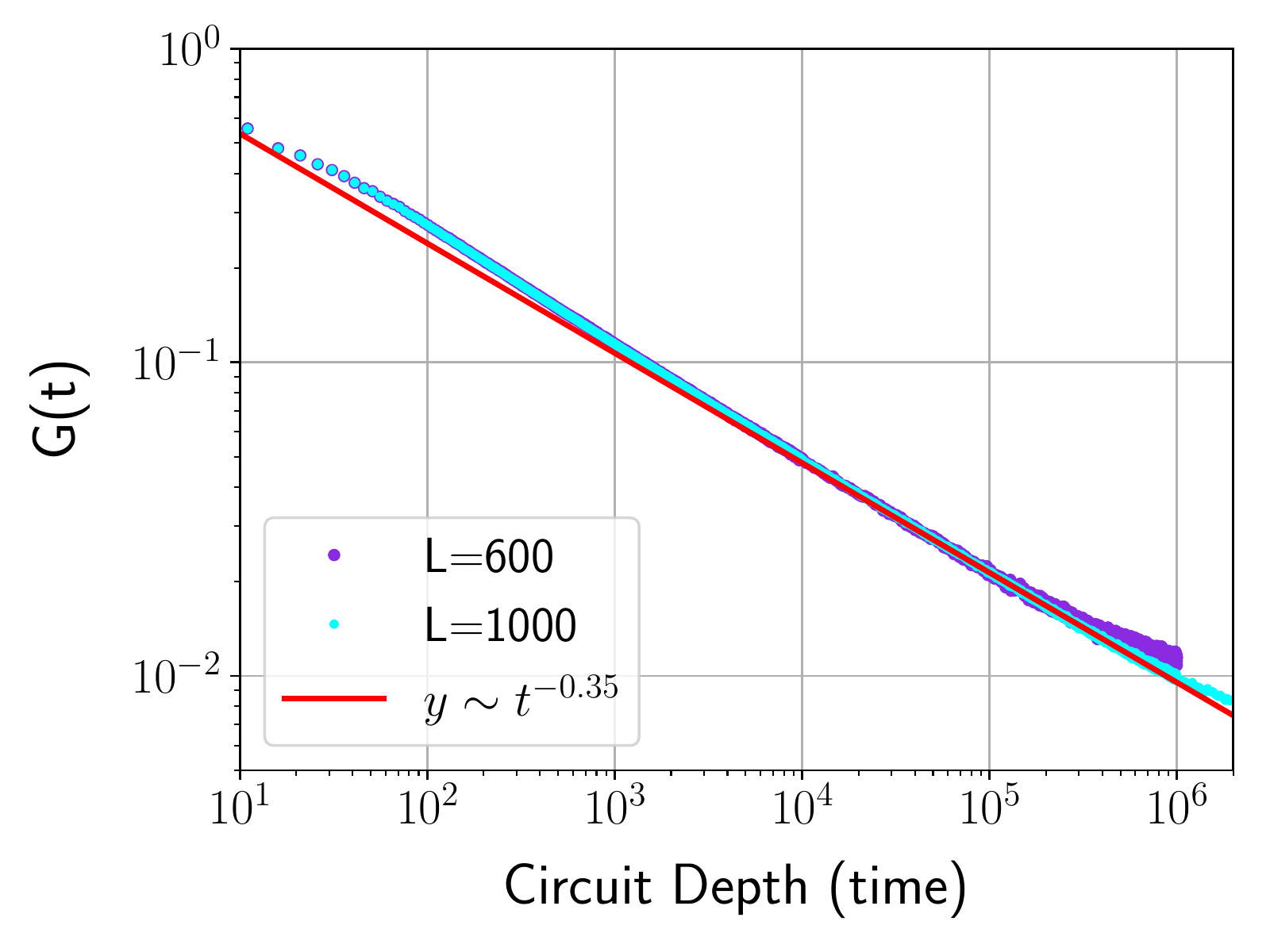}
\caption{The 2D triangular lattice autocorrelation function. {\it(top)} For the circuit with the smallest 
fundamental gates (3x3 gates), there are at least
two regimes. At relatively short times, the scaling is very slow,
$G(t) \sim t^{-\alpha}$ with $\alpha \ll \frac{1}{3}$. The scaling then slows
down at very long times, and in this regime it appears to approach the predicted
value $G(t) \sim t^{-1/3}$. {\it(bottom)} When gates of dimension 4x4 are used,
the relaxation to the power law scaling regime is much faster. Here, we see a
fit at late times to the form $G(t) \sim t^{-0.350(2)}$. This exponent is
very close to the predicted value of $\frac{1}{3}$. We expect that for even
larger systems and later times, the power law decay will converge to this
predicted value.} 
\label{fig:tri}
\end{figure}

We expect that for the triangular lattice, \cite{gromov2020}
\begin{eqnarray}
\partial_t G(x,t) = -\lambda \partial^2_{a_1} \partial^2_{a_2} \partial^2_{a_3}
G(x,t).
\end{eqnarray}
where $\partial_{a_i}$ denotes the derivative along the i-th lattice direction and $G(x,t) = \langle S^z(x,t)S^z(0,0)\rangle$.   In general, if there are $n$ linear constraints we might find
\begin{eqnarray}
\partial_t G(x,t) = (-1)^{n+1} \lambda \partial^2_{a_1} \partial^2_{a_2} \dots \partial^2_{a_n}
G(x,t).
\end{eqnarray}
Going back to the triangular lattice case, we expect that the real space autocorrelation function is 
then 
\begin{align}
G(r,t) &\sim \int d^2 k e^{i k r - \lambda k_{a_1}^2 k_{a_2}^2k_{a_3}^2 t}  \\
&= \int d^2 k^\prime e^{i
k r\cos\theta  - \lambda k^6 t \sin^2 (3\theta)} .
\end{align}
In the second equation we have switched to polar coordinates in wave number space.  Setting $r=0$ for simplicity, we find 
\begin{eqnarray}
G(0,t) \sim (\lambda t)^{-1/3} \int \frac{d\theta}{|\sin(3\theta)|^{2/3}} \sim t^{-1/3}.
\end{eqnarray} 
Note that on a square lattice where $\sin (3\theta)$ is replaced by $\sin(2\theta)$ and $k^6$ is replaced by $k^4$, there is a logarithmic correction coming from a divergent $\theta$ integral above. 

%Then, for $n$, constraints, we guess a solution of the form
%\begin{eqnarray}
%G(k,t) = e^{-\lambda t \prod_{i=1}^n k_i^2}
%\end{eqnarray}

%We note that for the previously studied orthogonal case, this gives $G(t)
%\sim t^{-\frac{1}{2}}$, which is correct up to the logarithmic correction found in \cite{IaconisVijay}. For the triangular lattice, we therefore expect that $G(t) \sim
%t^{-\frac{1}{3}}$.

%For this case, we look only at the autocorrelation function 
%\begin{eqnarray}
%G(t) = \langle S^z(t)S^z(0) \rangle.
%\end{eqnarray}
We now numerically calculate this autocorrelation function.  The results are shown in Fig. \ref{fig:tri}.  For gates of size $3\times 3$, the autocorrelation function takes a very long time to
reach the final scaling regime. Before this, the charge appears to go through a
regime where the decay is very slow.  The origin of this non-hydrodynamic effect is not understood.   For gates of size $4\times 4$, the decay rapidly approaches the hydrodynamic predictions above.
%Despite these issue, at the latest times, the decay rate does appear to approach
%$G(t) \sim t^-\frac{1}{3}$. Notice that we must go to very large lattice sizes
%and very long circuit depths to see this long wavelength behavior.  

\section{Discussion}
We have tested the analytic predictions of hydrodynamics \cite{gromov2020} for subdiffusion in multipole conserving systems numerically, using a numerical method from \cite{IaconisVijay}. Specifically, we have checked one dimensional systems with dipole and/or quadrupole conservation, two dimensional systems with dipole conservation, and triangular lattice systems with subsystem conservation laws along three non-orthogonal directions. In every case we find results in agreement with the analytic expectations. The numerical tests involve simulations of dynamics for very large system sizes and times. Our work thus demonstrates both the accuracy of the analytic predictions in \cite{gromov2020}, and the versatility and utility of the numerical method introduced in \cite{IaconisVijay}.  In the future, it would be interesting to extend these simulations to more exotic scenarios, perhaps with reduced spacetime symmetries, or with unconventional interplays between conserved multipoles and other conservation laws such as energy.

\section*{Acknowledgements}
 We acknowledge prior collaborations on related work with Andrey Gromov and Sagar Vijay. This material is based upon work supported in part (J.I. and R.N.) by the Air Force Office of Scientific Research under award number FA9550-20-1-0222.  A.L. was supported by a Research Fellowship from the Alfred P. Sloan Foundation.

\bibliography{base}

%merlin.mbs apsrev4-1.bst 2010-07-25 4.21a (PWD, AO, DPC) hacked
%Control: key (0)
%Control: author (0) dotless jnrlst
%Control: editor formatted (1) identically to author
%Control: production of article title (0) allowed
%Control: page (1) range
%Control: year (0) verbatim
%Control: production of eprint (0) enabled
\begin{thebibliography}{29}%
\makeatletter
\providecommand \@ifxundefined [1]{%
 \@ifx{#1\undefined}
}%
\providecommand \@ifnum [1]{%
 \ifnum #1\expandafter \@firstoftwo
 \else \expandafter \@secondoftwo
 \fi
}%
\providecommand \@ifx [1]{%
 \ifx #1\expandafter \@firstoftwo
 \else \expandafter \@secondoftwo
 \fi
}%
\providecommand \natexlab [1]{#1}%
\providecommand \enquote  [1]{``#1''}%
\providecommand \bibnamefont  [1]{#1}%
\providecommand \bibfnamefont [1]{#1}%
\providecommand \citenamefont [1]{#1}%
\providecommand \href@noop [0]{\@secondoftwo}%
\providecommand \href [0]{\begingroup \@sanitize@url \@href}%
\providecommand \@href[1]{\@@startlink{#1}\@@href}%
\providecommand \@@href[1]{\endgroup#1\@@endlink}%
\providecommand \@sanitize@url [0]{\catcode `\\12\catcode `\$12\catcode
  `\&12\catcode `\#12\catcode `\^12\catcode `\_12\catcode `\%12\relax}%
\providecommand \@@startlink[1]{}%
\providecommand \@@endlink[0]{}%
\providecommand \url  [0]{\begingroup\@sanitize@url \@url }%
\providecommand \@url [1]{\endgroup\@href {#1}{\urlprefix }}%
\providecommand \urlprefix  [0]{URL }%
\providecommand \Eprint [0]{\href }%
\providecommand \doibase [0]{http://dx.doi.org/}%
\providecommand \selectlanguage [0]{\@gobble}%
\providecommand \bibinfo  [0]{\@secondoftwo}%
\providecommand \bibfield  [0]{\@secondoftwo}%
\providecommand \translation [1]{[#1]}%
\providecommand \BibitemOpen [0]{}%
\providecommand \bibitemStop [0]{}%
\providecommand \bibitemNoStop [0]{.\EOS\space}%
\providecommand \EOS [0]{\spacefactor3000\relax}%
\providecommand \BibitemShut  [1]{\csname bibitem#1\endcsname}%
\let\auto@bib@innerbib\@empty
%</preamble>
\bibitem [{\citenamefont {Rigol}\ \emph {et~al.}(2008)\citenamefont {Rigol},
  \citenamefont {Dunjko},\ and\ \citenamefont {Olshanii}}]{Rigol}%
  \BibitemOpen
  \bibfield  {author} {\bibinfo {author} {\bibfnamefont {Marcos}\ \bibnamefont
  {Rigol}}, \bibinfo {author} {\bibfnamefont {Vanja}\ \bibnamefont {Dunjko}}, \
  and\ \bibinfo {author} {\bibfnamefont {Maxim}\ \bibnamefont {Olshanii}},\
  }\bibfield  {title} {\enquote {\bibinfo {title} {Thermalization and its
  mechanism for generic isolated quantum systems},}\ }\href
  {http://dx.doi.org/10.1038/nature06838} {\bibfield  {journal} {\bibinfo
  {journal} {Nature}\ }\textbf {\bibinfo {volume} {452}},\ \bibinfo {pages}
  {854--858} (\bibinfo {year} {2008})}\BibitemShut {NoStop}%
\bibitem [{\citenamefont {Nandkishore}\ and\ \citenamefont
  {Huse}(2015)}]{mblarcmp}%
  \BibitemOpen
  \bibfield  {author} {\bibinfo {author} {\bibfnamefont {Rahul}\ \bibnamefont
  {Nandkishore}}\ and\ \bibinfo {author} {\bibfnamefont {David~A.}\
  \bibnamefont {Huse}},\ }\bibfield  {title} {\enquote {\bibinfo {title}
  {Many-body localization and thermalization in quantum statistical
  mechanics},}\ }\href {\doibase 10.1146/annurev-conmatphys-031214-014726}
  {\bibfield  {journal} {\bibinfo  {journal} {Annual Review of Condensed Matter
  Physics}\ }\textbf {\bibinfo {volume} {6}},\ \bibinfo {pages} {15--38}
  (\bibinfo {year} {2015})},\ \Eprint
  {http://arxiv.org/abs/https://doi.org/10.1146/annurev-conmatphys-031214-014726}
  {https://doi.org/10.1146/annurev-conmatphys-031214-014726} \BibitemShut
  {NoStop}%
\bibitem [{\citenamefont {Nahum}\ \emph {et~al.}(2017)\citenamefont {Nahum},
  \citenamefont {Ruhman}, \citenamefont {Vijay},\ and\ \citenamefont
  {Haah}}]{Nahum1}%
  \BibitemOpen
  \bibfield  {author} {\bibinfo {author} {\bibfnamefont {Adam}\ \bibnamefont
  {Nahum}}, \bibinfo {author} {\bibfnamefont {Jonathan}\ \bibnamefont
  {Ruhman}}, \bibinfo {author} {\bibfnamefont {Sagar}\ \bibnamefont {Vijay}}, \
  and\ \bibinfo {author} {\bibfnamefont {Jeongwan}\ \bibnamefont {Haah}},\
  }\bibfield  {title} {\enquote {\bibinfo {title} {Quantum entanglement growth
  under random unitary dynamics},}\ }\href {\doibase 10.1103/PhysRevX.7.031016}
  {\bibfield  {journal} {\bibinfo  {journal} {Phys. Rev. X}\ }\textbf {\bibinfo
  {volume} {7}},\ \bibinfo {pages} {031016} (\bibinfo {year}
  {2017})}\BibitemShut {NoStop}%
\bibitem [{\citenamefont {Nahum}\ \emph {et~al.}(2018)\citenamefont {Nahum},
  \citenamefont {Vijay},\ and\ \citenamefont {Haah}}]{PhysRevX.8.021014}%
  \BibitemOpen
  \bibfield  {author} {\bibinfo {author} {\bibfnamefont {Adam}\ \bibnamefont
  {Nahum}}, \bibinfo {author} {\bibfnamefont {Sagar}\ \bibnamefont {Vijay}}, \
  and\ \bibinfo {author} {\bibfnamefont {Jeongwan}\ \bibnamefont {Haah}},\
  }\bibfield  {title} {\enquote {\bibinfo {title} {Operator spreading in random
  unitary circuits},}\ }\href {\doibase 10.1103/PhysRevX.8.021014} {\bibfield
  {journal} {\bibinfo  {journal} {Phys. Rev. X}\ }\textbf {\bibinfo {volume}
  {8}},\ \bibinfo {pages} {021014} (\bibinfo {year} {2018})}\BibitemShut
  {NoStop}%
\bibitem [{\citenamefont {{Nahum}}\ \emph {et~al.}(2018)\citenamefont
  {{Nahum}}, \citenamefont {{Ruhman}},\ and\ \citenamefont {{Huse}}}]{Nahum3}%
  \BibitemOpen
  \bibfield  {author} {\bibinfo {author} {\bibfnamefont {A.}~\bibnamefont
  {{Nahum}}}, \bibinfo {author} {\bibfnamefont {J.}~\bibnamefont {{Ruhman}}}, \
  and\ \bibinfo {author} {\bibfnamefont {D.~A.}\ \bibnamefont {{Huse}}},\
  }\bibfield  {title} {\enquote {\bibinfo {title} {{Dynamics of entanglement
  and transport in 1D systems with quenched randomness}},}\ }\href@noop {}
  {\bibfield  {journal} {\bibinfo  {journal} {Phys. Rev. B}\ }\textbf {\bibinfo
  {volume} {98}},\ \bibinfo {pages} {035118} (\bibinfo {year}
  {2018})}\BibitemShut {NoStop}%
\bibitem [{\citenamefont {{Chan}}\ \emph {et~al.}(2017)\citenamefont {{Chan}},
  \citenamefont {{De Luca}},\ and\ \citenamefont {{Chalker}}}]{amos1}%
  \BibitemOpen
  \bibfield  {author} {\bibinfo {author} {\bibfnamefont {A.}~\bibnamefont
  {{Chan}}}, \bibinfo {author} {\bibfnamefont {A.}~\bibnamefont {{De Luca}}}, \
  and\ \bibinfo {author} {\bibfnamefont {J.~T.}\ \bibnamefont {{Chalker}}},\
  }\bibfield  {title} {\enquote {\bibinfo {title} {{Solution of a minimal model
  for many-body quantum chaos}},}\ }\href@noop {} {\bibfield  {journal}
  {\bibinfo  {journal} {ArXiv e-prints}\ } (\bibinfo {year} {2017})},\ \Eprint
  {http://arxiv.org/abs/1712.06836} {arXiv:1712.06836 [cond-mat.stat-mech]}
  \BibitemShut {NoStop}%
\bibitem [{\citenamefont {{Chan}}\ \emph {et~al.}(2018)\citenamefont {{Chan}},
  \citenamefont {{De Luca}},\ and\ \citenamefont {{Chalker}}}]{amos2}%
  \BibitemOpen
  \bibfield  {author} {\bibinfo {author} {\bibfnamefont {A.}~\bibnamefont
  {{Chan}}}, \bibinfo {author} {\bibfnamefont {A.}~\bibnamefont {{De Luca}}}, \
  and\ \bibinfo {author} {\bibfnamefont {J.~T.}\ \bibnamefont {{Chalker}}},\
  }\bibfield  {title} {\enquote {\bibinfo {title} {{Spectral statistics in
  spatially extended chaotic quantum many-body systems}},}\ }\href@noop {}
  {\bibfield  {journal} {\bibinfo  {journal} {ArXiv e-prints}\ } (\bibinfo
  {year} {2018})},\ \Eprint {http://arxiv.org/abs/1803.03841} {arXiv:1803.03841
  [cond-mat.stat-mech]} \BibitemShut {NoStop}%
\bibitem [{\citenamefont {Kos}\ \emph {et~al.}(2018)\citenamefont {Kos},
  \citenamefont {Ljubotina},\ and\ \citenamefont {Prosen}}]{Prosen}%
  \BibitemOpen
  \bibfield  {author} {\bibinfo {author} {\bibfnamefont {Pavel}\ \bibnamefont
  {Kos}}, \bibinfo {author} {\bibfnamefont {Marko}\ \bibnamefont {Ljubotina}},
  \ and\ \bibinfo {author} {\bibfnamefont {Toma\ifmmode
  \check{z}\else~\v{z}\fi{}}\ \bibnamefont {Prosen}},\ }\bibfield  {title}
  {\enquote {\bibinfo {title} {Many-body quantum chaos: Analytic connection to
  random matrix theory},}\ }\href {\doibase 10.1103/PhysRevX.8.021062}
  {\bibfield  {journal} {\bibinfo  {journal} {Phys. Rev. X}\ }\textbf {\bibinfo
  {volume} {8}},\ \bibinfo {pages} {021062} (\bibinfo {year}
  {2018})}\BibitemShut {NoStop}%
\bibitem [{\citenamefont {Khemani}\ \emph {et~al.}(2018)\citenamefont
  {Khemani}, \citenamefont {Vishwanath},\ and\ \citenamefont
  {Huse}}]{KhemaniVishHuse}%
  \BibitemOpen
  \bibfield  {author} {\bibinfo {author} {\bibfnamefont {Vedika}\ \bibnamefont
  {Khemani}}, \bibinfo {author} {\bibfnamefont {Ashvin}\ \bibnamefont
  {Vishwanath}}, \ and\ \bibinfo {author} {\bibfnamefont {David~A.}\
  \bibnamefont {Huse}},\ }\bibfield  {title} {\enquote {\bibinfo {title}
  {Operator spreading and the emergence of dissipative hydrodynamics under
  unitary evolution with conservation laws},}\ }\href {\doibase
  10.1103/PhysRevX.8.031057} {\bibfield  {journal} {\bibinfo  {journal} {Phys.
  Rev. X}\ }\textbf {\bibinfo {volume} {8}},\ \bibinfo {pages} {031057}
  (\bibinfo {year} {2018})}\BibitemShut {NoStop}%
\bibitem [{\citenamefont {Rakovszky}\ \emph {et~al.}(2018)\citenamefont
  {Rakovszky}, \citenamefont {Pollmann},\ and\ \citenamefont {von
  Keyserlingk}}]{Keyserlingk1}%
  \BibitemOpen
  \bibfield  {author} {\bibinfo {author} {\bibfnamefont {Tibor}\ \bibnamefont
  {Rakovszky}}, \bibinfo {author} {\bibfnamefont {Frank}\ \bibnamefont
  {Pollmann}}, \ and\ \bibinfo {author} {\bibfnamefont {C.~W.}\ \bibnamefont
  {von Keyserlingk}},\ }\bibfield  {title} {\enquote {\bibinfo {title}
  {Diffusive hydrodynamics of out-of-time-ordered correlators with charge
  conservation},}\ }\href {\doibase 10.1103/PhysRevX.8.031058} {\bibfield
  {journal} {\bibinfo  {journal} {Phys. Rev. X}\ }\textbf {\bibinfo {volume}
  {8}},\ \bibinfo {pages} {031058} (\bibinfo {year} {2018})}\BibitemShut
  {NoStop}%
\bibitem [{\citenamefont {Chamon}(2005)}]{chamon}%
  \BibitemOpen
  \bibfield  {author} {\bibinfo {author} {\bibfnamefont {Claudio}\ \bibnamefont
  {Chamon}},\ }\bibfield  {title} {\enquote {\bibinfo {title} {Quantum
  glassiness in strongly correlated clean systems: An example of topological
  overprotection},}\ }\href {\doibase 10.1103/PhysRevLett.94.040402} {\bibfield
   {journal} {\bibinfo  {journal} {Phys. Rev. Lett.}\ }\textbf {\bibinfo
  {volume} {94}},\ \bibinfo {pages} {040402} (\bibinfo {year}
  {2005})}\BibitemShut {NoStop}%
\bibitem [{\citenamefont {Haah}(2011)}]{haah}%
  \BibitemOpen
  \bibfield  {author} {\bibinfo {author} {\bibfnamefont {Jeongwan}\
  \bibnamefont {Haah}},\ }\bibfield  {title} {\enquote {\bibinfo {title} {Local
  stabilizer codes in three dimensions without string logical operators},}\
  }\href {\doibase 10.1103/PhysRevA.83.042330} {\bibfield  {journal} {\bibinfo
  {journal} {Phys. Rev. A}\ }\textbf {\bibinfo {volume} {83}},\ \bibinfo
  {pages} {042330} (\bibinfo {year} {2011})}\BibitemShut {NoStop}%
\bibitem [{\citenamefont {Vijay}\ \emph {et~al.}(2015)\citenamefont {Vijay},
  \citenamefont {Haah},\ and\ \citenamefont {Fu}}]{fracton1}%
  \BibitemOpen
  \bibfield  {author} {\bibinfo {author} {\bibfnamefont {Sagar}\ \bibnamefont
  {Vijay}}, \bibinfo {author} {\bibfnamefont {Jeongwan}\ \bibnamefont {Haah}},
  \ and\ \bibinfo {author} {\bibfnamefont {Liang}\ \bibnamefont {Fu}},\
  }\bibfield  {title} {\enquote {\bibinfo {title} {A new kind of topological
  quantum order: A dimensional hierarchy of quasiparticles built from
  stationary excitations},}\ }\href {\doibase 10.1103/PhysRevB.92.235136}
  {\bibfield  {journal} {\bibinfo  {journal} {Phys. Rev. B}\ }\textbf {\bibinfo
  {volume} {92}},\ \bibinfo {pages} {235136} (\bibinfo {year}
  {2015})}\BibitemShut {NoStop}%
\bibitem [{\citenamefont {Vijay}\ \emph {et~al.}(2016)\citenamefont {Vijay},
  \citenamefont {Haah},\ and\ \citenamefont {Fu}}]{fracton2}%
  \BibitemOpen
  \bibfield  {author} {\bibinfo {author} {\bibfnamefont {Sagar}\ \bibnamefont
  {Vijay}}, \bibinfo {author} {\bibfnamefont {Jeongwan}\ \bibnamefont {Haah}},
  \ and\ \bibinfo {author} {\bibfnamefont {Liang}\ \bibnamefont {Fu}},\
  }\bibfield  {title} {\enquote {\bibinfo {title} {Fracton topological order,
  generalized lattice gauge theory, and duality},}\ }\href {\doibase
  10.1103/PhysRevB.94.235157} {\bibfield  {journal} {\bibinfo  {journal} {Phys.
  Rev. B}\ }\textbf {\bibinfo {volume} {94}},\ \bibinfo {pages} {235157}
  (\bibinfo {year} {2016})}\BibitemShut {NoStop}%
\bibitem [{\citenamefont {Pretko}(2017{\natexlab{a}})}]{sub}%
  \BibitemOpen
  \bibfield  {author} {\bibinfo {author} {\bibfnamefont {Michael}\ \bibnamefont
  {Pretko}},\ }\bibfield  {title} {\enquote {\bibinfo {title} {Subdimensional
  particle structure of higher rank $u(1)$ spin liquids},}\ }\href {\doibase
  10.1103/PhysRevB.95.115139} {\bibfield  {journal} {\bibinfo  {journal} {Phys.
  Rev. B}\ }\textbf {\bibinfo {volume} {95}},\ \bibinfo {pages} {115139}
  (\bibinfo {year} {2017}{\natexlab{a}})}\BibitemShut {NoStop}%
\bibitem [{\citenamefont {Nandkishore}\ and\ \citenamefont
  {Hermele}(2019)}]{fractonarcmp}%
  \BibitemOpen
  \bibfield  {author} {\bibinfo {author} {\bibfnamefont {Rahul~M}\ \bibnamefont
  {Nandkishore}}\ and\ \bibinfo {author} {\bibfnamefont {Michael}\ \bibnamefont
  {Hermele}},\ }\bibfield  {title} {\enquote {\bibinfo {title} {Fractons},}\
  }\href@noop {} {\bibfield  {journal} {\bibinfo  {journal} {Annual Review of
  Condensed Matter Physics}\ }\textbf {\bibinfo {volume} {10}},\ \bibinfo
  {pages} {295--313} (\bibinfo {year} {2019})}\BibitemShut {NoStop}%
\bibitem [{\citenamefont {Khemani}\ \emph {et~al.}(2020)\citenamefont
  {Khemani}, \citenamefont {Hermele},\ and\ \citenamefont {Nandkishore}}]{KN}%
  \BibitemOpen
  \bibfield  {author} {\bibinfo {author} {\bibfnamefont {Vedika}\ \bibnamefont
  {Khemani}}, \bibinfo {author} {\bibfnamefont {Michael}\ \bibnamefont
  {Hermele}}, \ and\ \bibinfo {author} {\bibfnamefont {Rahul}\ \bibnamefont
  {Nandkishore}},\ }\bibfield  {title} {\enquote {\bibinfo {title}
  {Localization from hilbert space shattering: From theory to physical
  realizations},}\ }\href {\doibase 10.1103/PhysRevB.101.174204} {\bibfield
  {journal} {\bibinfo  {journal} {Phys. Rev. B}\ }\textbf {\bibinfo {volume}
  {101}},\ \bibinfo {pages} {174204} (\bibinfo {year} {2020})}\BibitemShut
  {NoStop}%
\bibitem [{\citenamefont {Gromov}\ \emph {et~al.}(2020)\citenamefont {Gromov},
  \citenamefont {Lucas},\ and\ \citenamefont {Nandkishore}}]{gromov2020}%
  \BibitemOpen
  \bibfield  {author} {\bibinfo {author} {\bibfnamefont {Andrey}\ \bibnamefont
  {Gromov}}, \bibinfo {author} {\bibfnamefont {Andrew}\ \bibnamefont {Lucas}},
  \ and\ \bibinfo {author} {\bibfnamefont {Rahul~M.}\ \bibnamefont
  {Nandkishore}},\ }\bibfield  {title} {\enquote {\bibinfo {title} {Fracton
  hydrodynamics},}\ }\href {\doibase 10.1103/PhysRevResearch.2.033124}
  {\bibfield  {journal} {\bibinfo  {journal} {Phys. Rev. Research}\ }\textbf
  {\bibinfo {volume} {2}},\ \bibinfo {pages} {033124} (\bibinfo {year}
  {2020})}\BibitemShut {NoStop}%
\bibitem [{\citenamefont {Pai}\ \emph {et~al.}(2019)\citenamefont {Pai},
  \citenamefont {Pretko},\ and\ \citenamefont
  {Nandkishore}}]{pai2018localization}%
  \BibitemOpen
  \bibfield  {author} {\bibinfo {author} {\bibfnamefont {Shriya}\ \bibnamefont
  {Pai}}, \bibinfo {author} {\bibfnamefont {Michael}\ \bibnamefont {Pretko}}, \
  and\ \bibinfo {author} {\bibfnamefont {Rahul~M.}\ \bibnamefont
  {Nandkishore}},\ }\bibfield  {title} {\enquote {\bibinfo {title}
  {Localization in fractonic random circuits},}\ }\href {\doibase
  10.1103/PhysRevX.9.021003} {\bibfield  {journal} {\bibinfo  {journal} {Phys.
  Rev. X}\ }\textbf {\bibinfo {volume} {9}},\ \bibinfo {pages} {021003}
  (\bibinfo {year} {2019})}\BibitemShut {NoStop}%
\bibitem [{\citenamefont {Sala}\ \emph {et~al.}(2020)\citenamefont {Sala},
  \citenamefont {Rakovszky}, \citenamefont {Verresen}, \citenamefont {Knap},\
  and\ \citenamefont {Pollmann}}]{Sala_2020}%
  \BibitemOpen
  \bibfield  {author} {\bibinfo {author} {\bibfnamefont {Pablo}\ \bibnamefont
  {Sala}}, \bibinfo {author} {\bibfnamefont {Tibor}\ \bibnamefont {Rakovszky}},
  \bibinfo {author} {\bibfnamefont {Ruben}\ \bibnamefont {Verresen}}, \bibinfo
  {author} {\bibfnamefont {Michael}\ \bibnamefont {Knap}}, \ and\ \bibinfo
  {author} {\bibfnamefont {Frank}\ \bibnamefont {Pollmann}},\ }\bibfield
  {title} {\enquote {\bibinfo {title} {Ergodicity breaking arising from hilbert
  space fragmentation in dipole-conserving hamiltonians},}\ }\href {\doibase
  10.1103/physrevx.10.011047} {\bibfield  {journal} {\bibinfo  {journal}
  {Physical Review X}\ }\textbf {\bibinfo {volume} {10}} (\bibinfo {year}
  {2020}),\ 10.1103/physrevx.10.011047}\BibitemShut {NoStop}%
\bibitem [{\citenamefont {Iaconis}\ \emph {et~al.}(2019)\citenamefont
  {Iaconis}, \citenamefont {Vijay},\ and\ \citenamefont
  {Nandkishore}}]{IaconisVijay}%
  \BibitemOpen
  \bibfield  {author} {\bibinfo {author} {\bibfnamefont {Jason}\ \bibnamefont
  {Iaconis}}, \bibinfo {author} {\bibfnamefont {Sagar}\ \bibnamefont {Vijay}},
  \ and\ \bibinfo {author} {\bibfnamefont {Rahul}\ \bibnamefont
  {Nandkishore}},\ }\bibfield  {title} {\enquote {\bibinfo {title} {Anomalous
  subdiffusion from subsystem symmetries},}\ }\href {\doibase
  10.1103/PhysRevB.100.214301} {\bibfield  {journal} {\bibinfo  {journal}
  {Phys. Rev. B}\ }\textbf {\bibinfo {volume} {100}},\ \bibinfo {pages}
  {214301} (\bibinfo {year} {2019})}\BibitemShut {NoStop}%
\bibitem [{\citenamefont {Feldmeier}\ \emph {et~al.}(2020)\citenamefont
  {Feldmeier}, \citenamefont {Sala}, \citenamefont {de~Tomasi}, \citenamefont
  {Pollmann},\ and\ \citenamefont {Knap}}]{feldmeier2020}%
  \BibitemOpen
  \bibfield  {author} {\bibinfo {author} {\bibfnamefont {Johannes}\
  \bibnamefont {Feldmeier}}, \bibinfo {author} {\bibfnamefont {Pablo}\
  \bibnamefont {Sala}}, \bibinfo {author} {\bibfnamefont {Giuseppe}\
  \bibnamefont {de~Tomasi}}, \bibinfo {author} {\bibfnamefont {Frank}\
  \bibnamefont {Pollmann}}, \ and\ \bibinfo {author} {\bibfnamefont {Michael}\
  \bibnamefont {Knap}},\ }\href@noop {} {\enquote {\bibinfo {title} {Anomalous
  diffusion in dipole- and higher-moment conserving systems},}\ } (\bibinfo
  {year} {2020}),\ \Eprint {http://arxiv.org/abs/2004.00635} {arXiv:2004.00635
  [cond-mat.str-el]} \BibitemShut {NoStop}%
\bibitem [{\citenamefont {Morningstar}\ \emph {et~al.}(2020)\citenamefont
  {Morningstar}, \citenamefont {Khemani},\ and\ \citenamefont {Huse}}]{KH}%
  \BibitemOpen
  \bibfield  {author} {\bibinfo {author} {\bibfnamefont {Alan}\ \bibnamefont
  {Morningstar}}, \bibinfo {author} {\bibfnamefont {Vedika}\ \bibnamefont
  {Khemani}}, \ and\ \bibinfo {author} {\bibfnamefont {David~A.}\ \bibnamefont
  {Huse}},\ }\bibfield  {title} {\enquote {\bibinfo {title} {Kinetically
  constrained freezing transition in a dipole-conserving system},}\ }\href
  {\doibase 10.1103/PhysRevB.101.214205} {\bibfield  {journal} {\bibinfo
  {journal} {Phys. Rev. B}\ }\textbf {\bibinfo {volume} {101}},\ \bibinfo
  {pages} {214205} (\bibinfo {year} {2020})}\BibitemShut {NoStop}%
\bibitem [{\citenamefont {Zhang}(2020)}]{Pengfei}%
  \BibitemOpen
  \bibfield  {author} {\bibinfo {author} {\bibfnamefont {Pengfei}\ \bibnamefont
  {Zhang}},\ }\bibfield  {title} {\enquote {\bibinfo {title} {Subdiffusion in
  strongly tilted lattice systems},}\ }\href {\doibase
  10.1103/physrevresearch.2.033129} {\bibfield  {journal} {\bibinfo  {journal}
  {Physical Review Research}\ }\textbf {\bibinfo {volume} {2}},\ \bibinfo
  {pages} {033129} (\bibinfo {year} {2020})}\BibitemShut {NoStop}%
\bibitem [{\citenamefont {{Ganesan, Koushik and Lucas,
  Andrew}}(2020)}]{Ganesan}%
  \BibitemOpen
  \bibfield  {author} {\bibinfo {author} {\bibnamefont {{Ganesan, Koushik and
  Lucas, Andrew}}},\ }\bibfield  {title} {\enquote {\bibinfo {title}
  {{Holographic subdiffusion}},}\ }\href@noop {} {\bibfield  {journal}
  {\bibinfo  {journal} {{ArXiv e-prints}}\ } (\bibinfo {year} {2020})},\
  \Eprint {http://arxiv.org/abs/2008.09638} {arXiv:2008.09638 [hep-th]}
  \BibitemShut {NoStop}%
\bibitem [{\citenamefont {Gopalakrishnan}\ and\ \citenamefont
  {Zakirov}(2018)}]{GopalakrishnanBahti}%
  \BibitemOpen
  \bibfield  {author} {\bibinfo {author} {\bibfnamefont {Sarang}\ \bibnamefont
  {Gopalakrishnan}}\ and\ \bibinfo {author} {\bibfnamefont {Bahti}\
  \bibnamefont {Zakirov}},\ }\bibfield  {title} {\enquote {\bibinfo {title}
  {Facilitated quantum cellular automata as simple models with non-thermal
  eigenstates and dynamics},}\ }\href@noop {} {\bibfield  {journal} {\bibinfo
  {journal} {Quantum Science and Technology}\ }\textbf {\bibinfo {volume}
  {3}},\ \bibinfo {pages} {044004} (\bibinfo {year} {2018})}\BibitemShut
  {NoStop}%
\bibitem [{\citenamefont {Alba}\ \emph {et~al.}(2019)\citenamefont {Alba},
  \citenamefont {Dubail},\ and\ \citenamefont {Medenjak}}]{Alba}%
  \BibitemOpen
  \bibfield  {author} {\bibinfo {author} {\bibfnamefont {V.}~\bibnamefont
  {Alba}}, \bibinfo {author} {\bibfnamefont {J.}~\bibnamefont {Dubail}}, \ and\
  \bibinfo {author} {\bibfnamefont {M.}~\bibnamefont {Medenjak}},\ }\bibfield
  {title} {\enquote {\bibinfo {title} {Operator entanglement in interacting
  integrable quantum systems: The case of the rule 54 chain},}\ }\href
  {\doibase 10.1103/PhysRevLett.122.250603} {\bibfield  {journal} {\bibinfo
  {journal} {Phys. Rev. Lett.}\ }\textbf {\bibinfo {volume} {122}},\ \bibinfo
  {pages} {250603} (\bibinfo {year} {2019})}\BibitemShut {NoStop}%
\bibitem [{\citenamefont {Iaconis}(2020)}]{iaconis2020quantum}%
  \BibitemOpen
  \bibfield  {author} {\bibinfo {author} {\bibfnamefont {Jason}\ \bibnamefont
  {Iaconis}},\ }\href@noop {} {\enquote {\bibinfo {title} {Quantum state
  complexity in computationally tractable quantum circuits},}\ } (\bibinfo
  {year} {2020}),\ \Eprint {http://arxiv.org/abs/2009.05512} {arXiv:2009.05512
  [quant-ph]} \BibitemShut {NoStop}%
\bibitem [{\citenamefont {Pretko}(2017{\natexlab{b}})}]{genem}%
  \BibitemOpen
  \bibfield  {author} {\bibinfo {author} {\bibfnamefont {Michael}\ \bibnamefont
  {Pretko}},\ }\bibfield  {title} {\enquote {\bibinfo {title} {Generalized
  electromagnetism of subdimensional particles: A spin liquid story},}\ }\href
  {\doibase 10.1103/PhysRevB.96.035119} {\bibfield  {journal} {\bibinfo
  {journal} {Phys. Rev. B}\ }\textbf {\bibinfo {volume} {96}},\ \bibinfo
  {pages} {035119} (\bibinfo {year} {2017}{\natexlab{b}})}\BibitemShut
  {NoStop}%
\end{thebibliography}%


\begin{thebibliography}{11}
\expandafter\ifx\csname natexlab\endcsname\relax\def\natexlab#1{#1}\fi
\expandafter\ifx\csname bibnamefont\endcsname\relax
  \def\bibnamefont#1{#1}\fi
\expandafter\ifx\csname bibfnamefont\endcsname\relax
  \def\bibfnamefont#1{#1}\fi
\expandafter\ifx\csname citenamefont\endcsname\relax
  \def\citenamefont#1{#1}\fi
\expandafter\ifx\csname url\endcsname\relax
  \def\url#1{\texttt{#1}}\fi
\expandafter\ifx\csname urlprefix\endcsname\relax\def\urlprefix{URL }\fi
\providecommand{\bibinfo}[2]{#2}
\providecommand{\eprint}[2][]{\url{#2}}

\bibitem[{\citenamefont{Iaconis et~al.}(2019)\citenamefont{Iaconis, Vijay, and
  Nandkishore}}]{IaconisVijay}
\bibinfo{author}{\bibfnamefont{J.}~\bibnamefont{Iaconis}},
  \bibinfo{author}{\bibfnamefont{S.}~\bibnamefont{Vijay}}, \bibnamefont{and}
  \bibinfo{author}{\bibfnamefont{R.}~\bibnamefont{Nandkishore}},
  \bibinfo{journal}{Phys. Rev. B} \textbf{\bibinfo{volume}{100}},
  \bibinfo{pages}{214301} (\bibinfo{year}{2019}),
  \urlprefix\url{https://link.aps.org/doi/10.1103/PhysRevB.100.214301}.

\bibitem[{\citenamefont{Gromov et~al.}(2020)\citenamefont{Gromov, Lucas, and
  Nandkishore}}]{gromov2020}
\bibinfo{author}{\bibfnamefont{A.}~\bibnamefont{Gromov}},
  \bibinfo{author}{\bibfnamefont{A.}~\bibnamefont{Lucas}}, \bibnamefont{and}
  \bibinfo{author}{\bibfnamefont{R.~M.} \bibnamefont{Nandkishore}},
  \bibinfo{journal}{Phys. Rev. Research} \textbf{\bibinfo{volume}{2}},
  \bibinfo{pages}{033124} (\bibinfo{year}{2020}),
  \urlprefix\url{https://link.aps.org/doi/10.1103/PhysRevResearch.2.033124}.

\bibitem[{\citenamefont{Feldmeier et~al.}(2020)\citenamefont{Feldmeier, Sala,
  de~Tomasi, Pollmann, and Knap}}]{feldmeier2020}
\bibinfo{author}{\bibfnamefont{J.}~\bibnamefont{Feldmeier}},
  \bibinfo{author}{\bibfnamefont{P.}~\bibnamefont{Sala}},
  \bibinfo{author}{\bibfnamefont{G.}~\bibnamefont{de~Tomasi}},
  \bibinfo{author}{\bibfnamefont{F.}~\bibnamefont{Pollmann}}, \bibnamefont{and}
  \bibinfo{author}{\bibfnamefont{M.}~\bibnamefont{Knap}},
  \emph{\bibinfo{title}{Anomalous diffusion in dipole- and higher-moment
  conserving systems}} (\bibinfo{year}{2020}), \eprint{2004.00635}.

\bibitem[{\citenamefont{Nahum et~al.}(2018)\citenamefont{Nahum, Vijay, and
  Haah}}]{PhysRevX.8.021014}
\bibinfo{author}{\bibfnamefont{A.}~\bibnamefont{Nahum}},
  \bibinfo{author}{\bibfnamefont{S.}~\bibnamefont{Vijay}}, \bibnamefont{and}
  \bibinfo{author}{\bibfnamefont{J.}~\bibnamefont{Haah}},
  \bibinfo{journal}{Phys. Rev. X} \textbf{\bibinfo{volume}{8}},
  \bibinfo{pages}{021014} (\bibinfo{year}{2018}),
  \urlprefix\url{https://link.aps.org/doi/10.1103/PhysRevX.8.021014}.

\bibitem[{\citenamefont{Khemani et~al.}(2018)\citenamefont{Khemani, Vishwanath,
  and Huse}}]{KhemaniVishHuse}
\bibinfo{author}{\bibfnamefont{V.}~\bibnamefont{Khemani}},
  \bibinfo{author}{\bibfnamefont{A.}~\bibnamefont{Vishwanath}},
  \bibnamefont{and} \bibinfo{author}{\bibfnamefont{D.~A.} \bibnamefont{Huse}},
  \bibinfo{journal}{Phys. Rev. X} \textbf{\bibinfo{volume}{8}},
  \bibinfo{pages}{031057} (\bibinfo{year}{2018}),
  \urlprefix\url{https://link.aps.org/doi/10.1103/PhysRevX.8.031057}.

\bibitem[{\citenamefont{Rakovszky et~al.}(2018)\citenamefont{Rakovszky,
  Pollmann, and von Keyserlingk}}]{Keyserlingk1}
\bibinfo{author}{\bibfnamefont{T.}~\bibnamefont{Rakovszky}},
  \bibinfo{author}{\bibfnamefont{F.}~\bibnamefont{Pollmann}}, \bibnamefont{and}
  \bibinfo{author}{\bibfnamefont{C.~W.} \bibnamefont{von Keyserlingk}},
  \bibinfo{journal}{Phys. Rev. X} \textbf{\bibinfo{volume}{8}},
  \bibinfo{pages}{031058} (\bibinfo{year}{2018}),
  \urlprefix\url{https://link.aps.org/doi/10.1103/PhysRevX.8.031058}.

\bibitem[{\citenamefont{Pai et~al.}(2019)\citenamefont{Pai, Pretko, and
  Nandkishore}}]{pai2018localization}
\bibinfo{author}{\bibfnamefont{S.}~\bibnamefont{Pai}},
  \bibinfo{author}{\bibfnamefont{M.}~\bibnamefont{Pretko}}, \bibnamefont{and}
  \bibinfo{author}{\bibfnamefont{R.~M.} \bibnamefont{Nandkishore}},
  \bibinfo{journal}{Phys. Rev. X} \textbf{\bibinfo{volume}{9}},
  \bibinfo{pages}{021003} (\bibinfo{year}{2019}),
  \urlprefix\url{https://link.aps.org/doi/10.1103/PhysRevX.9.021003}.

\bibitem[{\citenamefont{Sala et~al.}(2020)\citenamefont{Sala, Rakovszky,
  Verresen, Knap, and Pollmann}}]{Sala_2020}
\bibinfo{author}{\bibfnamefont{P.}~\bibnamefont{Sala}},
  \bibinfo{author}{\bibfnamefont{T.}~\bibnamefont{Rakovszky}},
  \bibinfo{author}{\bibfnamefont{R.}~\bibnamefont{Verresen}},
  \bibinfo{author}{\bibfnamefont{M.}~\bibnamefont{Knap}}, \bibnamefont{and}
  \bibinfo{author}{\bibfnamefont{F.}~\bibnamefont{Pollmann}},
  \bibinfo{journal}{Physical Review X} \textbf{\bibinfo{volume}{10}}
  (\bibinfo{year}{2020}), ISSN \bibinfo{issn}{2160-3308},
  \urlprefix\url{http://dx.doi.org/10.1103/PhysRevX.10.011047}.

\bibitem[{\citenamefont{Khemani et~al.}(2020)\citenamefont{Khemani, Hermele,
  and Nandkishore}}]{KN}
\bibinfo{author}{\bibfnamefont{V.}~\bibnamefont{Khemani}},
  \bibinfo{author}{\bibfnamefont{M.}~\bibnamefont{Hermele}}, \bibnamefont{and}
  \bibinfo{author}{\bibfnamefont{R.}~\bibnamefont{Nandkishore}},
  \bibinfo{journal}{Phys. Rev. B} \textbf{\bibinfo{volume}{101}},
  \bibinfo{pages}{174204} (\bibinfo{year}{2020}),
  \urlprefix\url{https://link.aps.org/doi/10.1103/PhysRevB.101.174204}.

\bibitem[{\citenamefont{Morningstar et~al.}(2020)\citenamefont{Morningstar,
  Khemani, and Huse}}]{KH}
\bibinfo{author}{\bibfnamefont{A.}~\bibnamefont{Morningstar}},
  \bibinfo{author}{\bibfnamefont{V.}~\bibnamefont{Khemani}}, \bibnamefont{and}
  \bibinfo{author}{\bibfnamefont{D.~A.} \bibnamefont{Huse}},
  \bibinfo{journal}{Phys. Rev. B} \textbf{\bibinfo{volume}{101}},
  \bibinfo{pages}{214205} (\bibinfo{year}{2020}),
  \urlprefix\url{https://link.aps.org/doi/10.1103/PhysRevB.101.214205}.

\bibitem[{\citenamefont{Gopalakrishnan and
  Zakirov}(2018)}]{GopalakrishnanBahti}
\bibinfo{author}{\bibfnamefont{S.}~\bibnamefont{Gopalakrishnan}}
  \bibnamefont{and} \bibinfo{author}{\bibfnamefont{B.}~\bibnamefont{Zakirov}},
  \bibinfo{journal}{Quantum Science and Technology}
  \textbf{\bibinfo{volume}{3}}, \bibinfo{pages}{044004} (\bibinfo{year}{2018}).

\end{thebibliography}

\end{document}